\documentclass[aps,prb,reprint,amsmath,amssymb,nobibnotes,superscriptaddress]{revtex4-2}

\usepackage{bm}
\usepackage{graphicx}
\usepackage[utf8]{inputenc}
\usepackage{amsmath}
\usepackage{physics}
\usepackage{times}

\usepackage{hyperref} 
\hypersetup{colorlinks, allcolors=blue}

\begin{document}
\title{Strong-coupling approach to temperature dependence of competing orders of superconductivity: Possible time-reversal symmetry breaking and nontrivial topology} 

\author{Chi Sun}
\affiliation{\mbox{Center for Quantum Spintronics, Department of Physics, Norwegian University of Science and Technology, NO-7491 Trondheim, Norway}}
\author{Kristian M{\ae}land}
\affiliation{\mbox{Center for Quantum Spintronics, Department of Physics, Norwegian University of Science and Technology, NO-7491 Trondheim, Norway}}
\author{Even Thingstad}
\affiliation{Department of Physics, University of Basel, Klingelbergstrasse 82, CH-4056 Basel, Switzerland}
\author{Asle Sudb{\o}}
\email[Corresponding author: ]{asle.sudbo@ntnu.no}
\affiliation{\mbox{Center for Quantum Spintronics, Department of Physics, Norwegian University of Science and Technology, NO-7491 Trondheim, Norway}}

\begin{abstract}
We use strong-coupling Eliashberg theory to study the competition of separate superconducting orders at low temperatures. Specifically, we study magnon-mediated superconductivity in a trilayer heterostructure with a thin normal metal between two antiferromagnetic insulators. Spin-triplet $p$-wave, spin-triplet $f$-wave, and spin-singlet $d$-wave superconducting gaps have been predicted to occur close to the critical temperature for the superconducting instability. The gap symmetry with the largest critical temperature depends on parameters in the model. We confirm that the same gap symmetries appear at any temperature below the critical temperature. 
Furthermore, we show that the temperature can affect the competition between the different superconducting orders. 
In addition, we consider time-reversal-symmetry-breaking, complex linear combinations of candidate pairings, such as chiral $p$-, $f$-, and $d$-wave gaps, as well as $p_x+if_y$-wave gaps. We find indications that some of these time-reversal-symmetry-breaking, nodeless gaps offer a greater condensation energy than the time-reversal symmetric gaps.
This indicates that superconducting states with spontaneously broken time-reversal symmetry and nontrivial topology may be preferred in this system.     
\end{abstract}

\maketitle

\section{Introduction}
Superconductivity is a remarkable macroscopic quantum phenomenon \cite{SC_book,Sigrist1991Apr,Moore2020Aug,Stewart2011Dec}, which has been the subject of fascination even to the broader public ever since its discovery \cite{onnes1911leiden, vanDelft2010Sep}. In addition to endowing the photon with a mass via the Anderson-Higgs mechanism \cite{Anderson1962,Englert1964,Higgs1964}, completely altering the magnetic response of superconductors compared to normal metals, it features the complete absence of electrical resistance below a critical temperature. This property is highly attractive in various applications requiring extremely low power dissipation, high-speed operation, and high sensitivity \cite{Maksimov2000Oct,Nishijima2013Sep,Hassenzahl2004Sep}. According to the weak-coupling Bardeen-Cooper-Schrieffer (BCS) theory \cite{BCS}, conventional superconductivity is mediated by phonons, which bind electrons into Cooper pairs.
Magnons are bosonic quasiparticles describing spin fluctuations in magnets \cite{Chumak2015Jun}, and can replace phonons as the mediator of electron pairing to form superconductivity \cite{Brataas2020Nov}. Spin fluctuations close to an antiferromagnetic phase are believed to play an important role in cuprate and pnictide high-$T_c$ superconductors \cite{Pines1993, SCspinHistory_Scalapino1999, SCAFMspinMoriya2003Jul, SCspinGapSym_Hirschfeld2011}, as well as in Ce- and U-based heavy fermion superconductors \cite{HeavyFermionReview1984,NatRevMater2016}, possibly including chiral $p$-wave superconductors \cite{NatureChiralSC2020}. Furthermore, replacing phonons with magnons opens new routes for controlling unconventional forms of superconductivity.

Magnon-mediated superconductivity has also been proposed in heterostructures comprising magnetic insulators (MIs) and normal metals (NMs) \cite{Rohling2018Mar,Fjaerbu2019Sep,Erlandsen2019Sep,Eliash_Even,Sun2023Aug, Maeland2023Apr, Maeland2023Dec, Bostrom2023Dec}, where the MI/NM interface plays a crucial role in generating the effective attractive interaction between the electrons in the NM mediated by the magnons in the adjacent MI. 
The experiments \cite{Kajiwara2010NMMIexp, Li2016NMMIexp, Wu2016NMMIexp} have observed signatures of the interfacial exchange interaction between spins across NM/MI interfaces.
Additionally, Refs.~\cite{KargarianFMTI, Hugdal2018May, Hugdal2020FMTI, Erlandsen2020Mar} consider magnon-mediated superconductivity on the surface of a topological insulator.  
Both ferromagnetic insulators (FMIs) \cite{Rohling2018Mar} and antiferromagnetic insulators (AFMIs) \cite{Sun2023Aug,Eliash_Even,Erlandsen2019Sep,Fjaerbu2019Sep} have been considered for magnon-mediated superconductivity when attached with NMs. Compared with FMIs, AFMIs with compensated magnetic moments have the advantages of higher stability, lower energy dissipation, and faster magnetization dynamics \cite{Jungwirth2016Mar,Sun2022Mar}. 
Moreover, the two sublattices in the AFMI can couple asymmetrically to the electrons in the NM at the AFMI/NM interface~\cite{Nogues2005, Wu2020}. This asymmetry can be achieved by modulating the exposure of the two sublattices to the interface, providing an additional adjustable parameter to tune and optimize the magnon-mediated superconductivity \cite{Kamra2019afmsqueezed, Erlandsen2019Sep, Eliash_Even, Sun2023Aug}. 

In an AFMI/NM heterostructure with matching square lattices, the spatial periodicity in the magnet is larger than in the electronic system. Consequently, the electron first Brillouin zone (1BZ) is twice as big as the magnon Brillouin zone, which we therefore refer to as the reduced Brillouin zone (RBZ).  
This introduces electron-magnon scattering of two types, namely regular and Umklapp processes \cite{Eliash_Even, Sun2023Aug}. For a NM with small filling and a small Fermi surface (FS), the electrons close to the FS are scattered with momenta within the magnon RBZ through the regular processes. 
Approaching half filling, the FS grows, and this allows Umklapp scattering processes involving momenta outside the RBZ. 
Depending on the filling and the sublattice coupling asymmetry, the interplay of regular and Umklapp processes results in different magnon-mediated superconducting phases \cite{Erlandsen2019Sep,Eliash_Even,Sun2023Aug}. For significant sublattice coupling asymmetry, small filling and small FSs results in a $p$-wave phase, while an $f$-wave phase is obtained for larger FSs when approaching half-filling. Close to sublattice symmetry, a $d$-wave pairing is preferred close to half-filling. 
This was shown using a strong-coupling approach in Ref.~\cite{Eliash_Even} at the critical temperature, and using a weak-coupling approach at the critical and at zero temperature in Ref.~\cite{Sun2023Aug}.
Thus, the phase diagrams have only been discussed at specific temperatures.  
It is of interest to investigate the temperature dependence of the superconducting gap function in the entire temperature regime below the critical temperature, since mixed states may become preferred \cite{Hutchinson2020Nov}.

In this paper, we investigate magnon-mediated superconductivity in a NM induced by interfacial coupling to AFMIs at any nonzero temperature within a strong-coupling Eliashberg theory framework \cite{Eliashberg1960Sep, Eliashberg1961, Carbotte1990FreeEnergy, EliashbergRevMarsiglio2020, Chubukov2020FEspecialized}. We investigate the competition between superconducting gaps with different gap symmetries, including complex linear combinations. Gaps with a phase difference between different pairing channels break the time-reversal symmetry (TRS) of the model Hamiltonian, and provide nodeless gaps. 
The spontaneous breaking of symmetries can lead to qualitatively new effects in unconventional superconductors~\cite{Sigrist1991Apr, Sigrist2000TRSB, Fernandes2022FeSC, Aase2023Dec, Shang2018ReExp, Csire2022ReTheory}.
For instance, chiral $p$-, $d$-, and $f$-wave superconductors are topological and have received significant attention due to interesting conducting channels at interfaces and possible applications in topological quantum computation \cite{Huang2014chiralpdf, Tada2015chiralpdf, Volovik2015chiralpdf, Suzuki2023chiralpdf, Black-Schaffer2014dwave, Serban2010chiralpdomain, Lundemo2024Jan, Maeland2023Apr, Maeland2023Dec, Bostrom2023Dec, Schnyder2008tenfold, Bernevig2013,  TopoSCrevSato, Leijnse2012TSCrev, TopoQuantumCompRevModPhys}.
Additionally, TRS breaking has been studied in multiband \cite{Fernandes2022FeSC, Aase2023Dec} and multiorbital \cite{Shang2018ReExp, Csire2022ReTheory} superconductors, where magnetic fluctuations could play a role \cite{Fernandes2022FeSC, Aase2023Dec, Csire2022ReTheory}.

Eliashberg theory takes many-body renormalization of the electron bands into account, which could be significant for magnon-mediated interactions \cite{KargarianFMTI, Maeland2021Sep, Eliash_Even}. Furthermore, Eliashberg theory captures retardation by considering the frequency dependence of the magnon-mediated interaction~\cite{Eliashberg1960Sep, Eliashberg1961, Carbotte1990FreeEnergy, EliashbergRevMarsiglio2020, Chubukov2020FEspecialized}.
Together, these effects give significant corrections to the critical temperature obtained through BCS theory \cite{Eliash_Even}. By extension, they should also affect the superconducting gap amplitude, and this could influence the competition between the different superconducting gap symmetries relative to the BCS result at zero temperature~\cite{Sun2023Aug}.

Instead of solving the linearized Eliashberg equations at the critical temperature~\cite{Eliash_Even},
we consider the nonlinear Eliashberg equations that yield the superconducting gap at any temperature. 
Taking both regular and Umklapp processes into account, we solve the Eliashberg equations on the FS numerically.
By adjusting the chemical potential in the NM and the asymmetry of the interfacial coupling to the two sublattices of the AFMIs, different superconducting gap symmetries ($p$-, $f$- and $d$-wave) are obtained at low temperatures, which is in agreement with the gap symmetries found at the critical temperature in Ref.~\cite{Eliash_Even}. 
Our approach allows exploration of how the development of finite gaps affects the location of the transitions between the superconducting phases. Finally, we explore how TRS-breaking complex linear combinations of candidate pairings may be energetically preferred over TRS-preserving gaps. This indicates the possibility for spinful topological superconductivity.

\section{Theory}
\subsection{Model}
\begin{figure}[ht]
  \centering
\includegraphics[width=0.4\columnwidth]{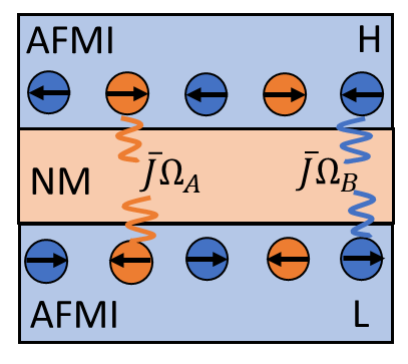}
  \caption{Schematic representation of the trilayer structure considered in this paper, with a normal metal (NM) sandwiched between two antiferromagnetic insulator (AFMI) layers on a bipartite lattice (orange and blue). The two AFMIs are oppositely oriented. The coupling to the same AFMI sublattice is taken to be the same at both interfaces. Meanwhile, the coupling strength can be different for the two sublattices.}%
  \label{fig:trilayer}
\end{figure}
We consider the AFMI/NM/AFMI trilayer structure shown in Fig.~\ref{fig:trilayer}, which is modeled by the Hamiltonian $H=H_\text{AFMI} + H_\text{NM} + H_\text{int}$. We assume that the NM and the AFMIs are sufficiently thin to be accurately described as two-dimensional (2D) monolayers, and consider Hamiltonians describing the AFMI,  the NM, and the interfacial interaction given by \cite{Eliash_Even}
\begin{align}
    H_\text{AFMI} = \sum_{i,j,\eta}J_{ij}\boldsymbol{S}_{i\eta}\cdot \boldsymbol{S}_{j\eta} - K\sum_{i,\eta}(S_{i\eta}^z)^2,\label{eq:H_AFMI}\\
    H_\text{NM} = -t\sum_{\langle i,j \rangle,\sigma}c_{i\sigma}^{\dag}c_{j\sigma} - \mu\sum_{i,\sigma}c_{i\sigma}^\dag c_{i\sigma},\label{eq:H_NM}\\
    H_\text{int}=-2\bar{J}\sum_{\eta,\Upsilon}\sum_{i\in \Upsilon}\Omega_\Upsilon^\eta c_i^\dag \boldsymbol{\sigma} c_i \cdot \boldsymbol{S}_{i\eta},
\end{align}
in which the sum over $i$, $j$ denotes the sum over lattice sites on a square lattice, the sum over $\eta \in \{H,L\}$ denotes the sum over the two AFMIs, and the sum over $\Upsilon \in \{A,B\}$ denotes the sum over the two AFMI sublattices. In the AFMI, $\boldsymbol{S}_i$ is the spin-$S$ operator at site $i$, $J_{ij}$ is the exchange coupling (between spins at sites $i$ and $j$), and $K$ is the easy-axis anisotropy (along $\hat{z}$). Here we take $J_{1(2)}$ for the (next-)nearest neighbors and set other $J_{ij}$ to zero. In the NM,  $c_{i\sigma}^\dag$ ($c_{i\sigma}$) is the electron creation (annihilation) operator, which creates (annihilates) an electron with spin $\sigma$ at site $i$. Furthermore, $t$ is the tight-binding hopping parameter for nearest neighbors and $\mu$ denotes the chemical potential. At the AFMI/NM interface, $\bar{J}$ describes the interfacial exchange coupling between the lattice spins in the AFMI and the conduction electron spins in the NM. Here a dimensionless, sublattice and layer-dependent, parameter $\Omega_{\Upsilon}^\eta$ is utilized to introduce asymmetry of the interfacial exchange coupling. In order to eliminate any magnetic fields in the NM, we consider equal coupling to the two AFMI layers, i.e., $\Omega_{\Upsilon}^\eta \equiv \Omega_{\Upsilon}$. The notation $c_i \equiv (c_{i\uparrow},c_{i\downarrow})^T$ is introduced and $\boldsymbol{\sigma}$ is the Pauli matrix vector. In addition, we set $\hbar=a=1$, with $a$ being the lattice constant. 

We now derive the Eliashberg equations for the system following Ref.~\cite{Eliash_Even}. 
To allow the possibility of complex gaps, we parametrize the self-energy by
\begin{align}
    \Sigma =& (1-Z)i\omega_n   \tau_0 \sigma_0 + \chi  \tau_3 \sigma_0 + \phi_s^R  \tau_2 \sigma_2 + \phi_s^I \tau_1 \sigma_2 \nonumber\\
    & + \phi_t^R \tau_1 \sigma_1 + \phi_t^I \tau_2 \sigma_1,
\end{align}
where $Z$ is the electron renormalization, $\chi$ is the quasiparticle energy shift, and $\phi_{s(t)}$ describes the spin singlet (triplet) superconducting pairing. 
Although explicit dependence is suppressed in the notation above, these fields depend on momentum and frequency. We use the notation $k=(\boldsymbol{k},i\omega_n)$ with momentum $\boldsymbol{k}=(k_x,k_y)$ and fermionic Matsubara frequency $\omega_n=(2n+1)\pi/\beta$, where $\beta$ is the inverse temperature.
Furthermore, we use the shorthand notation $\tau_i \sigma_j$ for the Pauli matrix outer product $\tau_i \otimes \sigma_j$. 
$\tau_i$ covers the particle-hole degree of freedom, while $\sigma_i$ acts on the spin degree of freedom. 
The off-diagonal terms in the particle-hole sector are related to anomalous Green's functions describing superconducting pairing. Note that the above self-energy only describes Cooper pairs of electrons with opposite spin since the pairing interaction does not allow spin-polarized pairs.
We now define complex spin singlet and spin triplet pairing functions through $\phi_{s,t} = \phi_{s,t}^R + i \phi_{s,t}^I$. 
Exploiting the symmetries of the Green's function~\cite{Eliash_Even}, we find $\phi_{s,t}(k) = - \zeta_{s,t} \phi_{s,t}(-k)$, where \(\zeta_s = -1\) and \(\zeta_t = 1\).
The Eliashberg equations of the trilayer system are 
\begin{align}
    [1-Z(k)]i\omega_n = -V^2\frac{1}{\beta}\sum_{k{'}}D(k-k{'})\frac{i\omega_{n{'}}Z(k{'})}{\Theta(k{'})},\label{Eli0_1}\\ 
    \phi_{s,t}(k) = \zeta_{s,t}V^2\frac{1}{\beta}\sum_{k{'}}D(k-k{'})\frac{\phi_{s,t}(k{'})}{\Theta(k{'})}\label{Eli0_2},
\end{align}
in which we have omitted the equation for the quasiparticle energy shift $\chi(k)$ 
since it only represents an integration variable shift when the perpendicular momentum is integrated out to obtain the Eliashberg equations on the FS.  
Furthermore, we assumed that the pairing function is either spin singlet or spin triplet, and their coexistence is not covered. 
The interaction strength parameter is defined as $V\equiv-2\bar{J}\sqrt{S/N}$ 
with $N$ being the number of lattice sites at the AFMI/NM interface. The denominator is given by
\begin{equation}
    \Theta(k)=[i\omega_n Z(k)]^2-\xi_{\boldsymbol{k}}^2-|\phi_{s,t}(k)|^2,
\end{equation}
in which $\xi_{\boldsymbol{k}}=-2t(\cos{k_x}+\cos{k_y})-\mu$ is the diagonalized dispersion relation of the NM described by Eq.~(\ref{eq:H_NM}) through a Fourier transform (FT). 
The magnon propagator $D$ is given by
\begin{equation}
    D(q)=\theta_{\boldsymbol{q}}D^{RR}(\boldsymbol{q},i\nu_m) + \theta_{\boldsymbol{q}+\boldsymbol{Q}}D^{UU}(\boldsymbol{q}+\boldsymbol{Q},i\nu_m),
\end{equation}
where $\boldsymbol{q}=\boldsymbol{k}-\boldsymbol{k}{'}$ can take the values in the full electron Brillouin zone by combing the regular (R) and Umklapp (U) scattering processes and $\boldsymbol{Q}=\pi(\hat{x} + \hat{y})$ is the AFM ordering vector. $\theta_{\boldsymbol{q}}$ is defined as 1 (0) when $\boldsymbol{q}$ resides inside (outside) the RBZ. $D^{RR/UU}$ is the regular/Umklapp noninteracting eigenmagnon propagator given by $D^{RR/UU}(\boldsymbol{q},i\nu_m)=-2A^{RR/UU}(\boldsymbol{q})2\omega_{\boldsymbol{q}}/(\nu_m^2+\omega_{\boldsymbol{q}}^2)$, in which $\nu_m=2m\pi/\beta$ is a bosonic Matsubara frequency and $\omega_{\boldsymbol{q}} = \sqrt{C_{\boldsymbol{q}}^2 - D_{\boldsymbol{q}}^2}$ is the AFMI magnon dispersion. Here $C_{\boldsymbol{q}}=2z_1J_1S-2z_2J_2S(1-\Tilde{\gamma}_{\boldsymbol{q}})+2KS$ and $D_{\boldsymbol{q}}=2z_1J_1S\gamma_{\boldsymbol{q}}$, in which the structure factors $\gamma_{\boldsymbol{q}}=\sum_{\boldsymbol{\delta}_1} e^{i\boldsymbol{q}\cdot \boldsymbol{\delta}_1}/z_1$ and $\Tilde{\gamma}_{\boldsymbol{q}}=\sum_{\boldsymbol{\delta}_2} e^{i\boldsymbol{q}\cdot \boldsymbol{\delta}_2}/z_2$ are defined for the (next-)nearest neighbor vector $\boldsymbol{\delta}_1$ ($\boldsymbol{\delta}_2$) with $z_1$ ($z_2$) as the corresponding number of neighbors.
The magnon bands are obtained from Eq.~(\ref{eq:H_AFMI}) by performing Holstein-Primakoff, Fourier, and Bogoliubov transformations with appropriate coherence factors $u_{\boldsymbol{q}}=\cosh{\tilde{\theta}_{\boldsymbol{q}}}$ and $v_{\boldsymbol{q}}=\sinh{\tilde{\theta}_{\boldsymbol{q}}}$ with $\tilde{\theta}_{\boldsymbol{q}}=-\operatorname{artanh}{(D_{\boldsymbol{q}}/C_{\boldsymbol{q}})}/2$. The boosting factors are, in terms of the coherence factors, given by $A^{RR/UU}(\boldsymbol{q})=[(\Omega_A u_{\boldsymbol{q}} \pm \Omega_B v_{\boldsymbol{q}})^2+(\Omega_A v_{\boldsymbol{q}} \pm \Omega_B u_{\boldsymbol{q}})^2]/2$. More details can be found in Ref.~\cite{Eliash_Even}.

\subsection{Fermi surface averaged Eliashberg equations}
Here we assume the momentum regions close to the FS dominate when performing momentum sums in the Eliashberg equations, which should be the case when the electron energy scale is much larger than the magnon energy scale. Further, we let $Z(k)=Z(i\omega_n)$ and $\phi_{s,t}(k)=\psi(\boldsymbol{k})\phi_{s,t}(i\omega_n)$ for momenta close to the FS.

To get the FS-averaged version of Eq.~(\ref{Eli0_1}), we consider the general form
\begin{equation}
f(k)=-V^2\frac{1}{\beta}\sum_{\boldsymbol{k}{'}}D(k-k{'})h(\boldsymbol{k}',\xi_{\boldsymbol{k}'}),  
\end{equation}
in which $f(k) = f(i\omega_n)$ and $h(\boldsymbol{k}',\xi_{\boldsymbol{k}'})=i\omega_{n{'}}Z(i\omega_{n{'}})/\Theta(k{'})$. By splitting $\boldsymbol{k}{'}$ into a perpendicular and a parallel part and converting the perpendicular momentum integral to an energy integral, the FS-averaged quantity becomes
\begin{align}
  f(i\omega_n)&=-V^2\frac{1}{\beta N_F}\sum_{\boldsymbol{k},\boldsymbol{k}{'}}\delta(\xi_{\boldsymbol{k}})\delta(\xi_{\boldsymbol{k}{'}})D(k-k{'})\notag\\&\times\bigg[\frac{1}{N_F}\int d\xi h(\boldsymbol{k}{'},\xi)N(\xi)\bigg],
\label{eq:fbar0_T}
\end{align}
in which $N_F=\sum_{\boldsymbol{k}}\delta(\xi_{\boldsymbol{k}})$ is the density of states per spin at the FS. The integral in the square bracket in Eq.~\eqref{eq:fbar0_T} can be calculated as
\begin{align}
&\frac{1}{N_F}\int d\xi h(\boldsymbol{k}{'},\xi)N(\xi)\approx \int_{-\infty}^{\infty} d\xi h(\boldsymbol{k}{'},\xi)\notag\\
&=\int_{-\infty}^{\infty} d\xi \frac{i\omega_{n{'}}Z(i\omega_{n{'}})}{[i\omega_{n{'}} Z(i\omega_{n{'}})]^2-\xi^2-|\psi(\boldsymbol{k}{'})\phi_{s,t}(i\omega_{n{'}})|^2}\notag\\
&=\frac{-\pi i\omega_{n{'}}Z(i\omega_{n{'}})}{\sqrt{[\omega_{n{'}} Z(i\omega_{n{'}})]^2+|\psi(\boldsymbol{k}{'})\phi_{s,t}(i\omega_{n{'}})|^2}},
\label{eq:2nd0_bracket_T}
\end{align}
in which we have utilized $\int_{-\infty}^{+\infty}dx/(x^2+a^2)=\pi/a$ to integrate out the energy. 
Inserting Eq.~(\ref{eq:2nd0_bracket_T}) into Eq.~(\ref{eq:fbar0_T}) and including the sum over $\omega_{n{'}}$, the FS-averaged version of Eq.~(\ref{Eli0_1}) is given by
\begin{align}
   [1-Z(i\omega_n)]i\omega_n=\frac{iV^2\pi}{\beta N_F}\sum_{\boldsymbol{k},\boldsymbol{k}{'},\omega_{n{'}}}\delta(\xi_{\boldsymbol{k}})\delta(\xi_{\boldsymbol{k}{'}})\notag\\\times\frac{D(k-k{'})\omega_{n{'}}Z(i\omega_{n{'}})}{\sqrt{[\omega_{n{'}} Z(i\omega_{n{'}})]^2+|\psi(\boldsymbol{k}{'})\phi_{s,t}(i\omega_{n{'}})|^2}}. 
\end{align}

Following the same procedure for $f(k) = f(i\omega_n)\psi(\boldsymbol{k})$ and $h(\boldsymbol{k}',\xi_{\boldsymbol{k}{'}})=\phi_{s,t}(k{'})/\Theta(k{'})$, the FS-averaged version of Eq.~(\ref{Eli0_2}) becomes
\begin{align}
    \phi_{s,t}(i\omega_n)=\frac{-\zeta_{s,t}V^2}{\langle |\psi(\boldsymbol{k})|^2 \rangle_{\text{FS}}}\frac{\pi}{\beta N_F}\sum_{\boldsymbol{k},\boldsymbol{k}{'},\omega_{n{'}}}\delta(\xi_{\boldsymbol{k}})\delta(\xi_{\boldsymbol{k}{'}})\notag\\\times \frac{\psi^*(\boldsymbol{k})D(k-k{'})\psi(\boldsymbol{k}{'})\phi_{s,t}(i\omega_{n{'}})}{\sqrt{[\omega_{n{'}} Z(i\omega_{n{'}})]^2+|\psi(\boldsymbol{k}{'})\phi_{s,t}(i\omega_{n{'}})|^2}},
\end{align}
in which $\langle \cdots \rangle_{\text{FS}}$ denotes the FS average $\langle f(\boldsymbol{k}) \rangle_{\text{FS}} =  \sum_{\boldsymbol{k}} \delta(\xi_{\boldsymbol{k}}) f(\boldsymbol{k})/N_F.$

Introducing the superconducting gap function $\Delta_{s,t}(k)=\phi_{s,t}(k)/Z(k)$, we rewrite the Eliashberg equations as
\begin{align}
   Z(i\omega_n)&=1-V^2\frac{\pi}{\beta N_F \omega_n}\sum_{\boldsymbol{k},\boldsymbol{k}{'},\omega_{n{'}}}\delta(\xi_{\boldsymbol{k}})\delta(\xi_{\boldsymbol{k}{'}})D(k-k{'})\notag\\&\times\frac{\omega_{n{'}}}{\sqrt{\omega_{n{'}}^2+|\psi(\boldsymbol{k}{'})\Delta_{s,t}(i\omega_{n{'}})|^2}},
\label{eq:1}
\end{align}
\begin{align}
   \Delta_{s,t}(i\omega_n)&=\frac{-\zeta_{s,t}V^2}{\langle |\psi(\boldsymbol{k})|^2 \rangle_{\text{FS}} Z(i\omega_n)}\frac{\pi}{\beta N_F}\sum_{\boldsymbol{k},\boldsymbol{k}{'},\omega_{n{'}}}\delta(\xi_{\boldsymbol{k}})\notag\\&\times\delta(\xi_{\boldsymbol{k}{'}})\frac{\psi^*(\boldsymbol{k})D(k-k{'})\psi(\boldsymbol{k}{'})\Delta_{s,t}(i\omega_{n{'}})}{\sqrt{\omega_{n{'}}^2+|\psi(\boldsymbol{k}{'})\Delta_{s,t}(i\omega_{n{'}})|^2}}.
\label{eq:2}
\end{align}
In contrast to the linearized Eliashberg equations at the critical temperature \cite{Eliash_Even}, Eqs.~(\ref{eq:1}) and (\ref{eq:2}) are coupled and nonlinear, describing the electron band renormalization and superconducting gap at any nonzero temperature. They are solved by fixed-point iteration, and the sums over momenta must be calculated at each step in the iteration. We introduce a symmetric cutoff in the sums over Matsubara frequencies. The cutoff is increased until the results converge. Since large numbers of Matsubara frequencies are needed at low temperatures, the fast Fourier transform is implemented to speed up the calculation of the convolution in the Eliashberg equations, i.e., 
\begin{align}
    \sum_{i\omega_{n'}} D(i\omega_n-i\omega_{n'}) F(i\omega_{n'}) =& \mathcal{F}^{-1}\big([\mathcal{F}D(i\omega_n-i\omega_{n'})]\nonumber\\
    &\times[\mathcal{F}F(i\omega_{n'})]\big),
\end{align}
where \(\mathcal{F}\) denotes the temporal Fourier transform of some frequency dependent function $F$ and $\boldsymbol{k}$ and $\boldsymbol{k}'$ dependencies are omitted. 

\section{results and discussion}
To tune the sublattice coupling asymmetry, we set $\Omega_A=1$ and $\Omega_B=\Omega\in[0,1]$. We substitute different ansatze for $\psi(\boldsymbol{k})$ to investigate different superconducting gap symmetries. In particular, we consider three ansatze: spin-triplet $p$-wave, spin-triplet $f$-wave, and spin-singlet $d$-wave described by
\begin{align}
\psi_{p_x}(\boldsymbol{k}) &= \cos{\phi_{\boldsymbol{k}}}, &  \psi_{p_y}(\boldsymbol{k}) &= \sin{\phi_{\boldsymbol{k}}},\label{eq:pwave}\\
\psi_{d_{x^2-y^2}}(\boldsymbol{k}) &= \cos{2\phi_{\boldsymbol{k}}},
& \psi_{d_{xy}}(\boldsymbol{k}) &= \sin{2\phi_{\boldsymbol{k}}},\label{eq:dwave} \\
\psi_{f_x}(\boldsymbol{k}) &= \cos{3\phi_{\boldsymbol{k}}}, &  \psi_{f_y}(\boldsymbol{k}) &= \sin{3\phi_{\boldsymbol{k}}},\label{eq:fwave}
\end{align}
in which $\phi_{\boldsymbol{k}} = \arg (k_x + i k_y)$ is the polar angle between $\boldsymbol{k}$ on the FS and the $x$ axis. Combinations of the above gap symmetries are also considered. 
Note that the $f$-wave functions are cubic, with six zeros on a circle around the origin. $f_x$ is a shorthand for $f_{x(x^2-3y^2)} \sim k_x(k_x^2-3k_y^2)$ and $f_y$ is a shorthand for $f_{y(3x^2-y^2)} \sim k_y(3k_x^2-k_y^2)$ \cite{Maeland2023Apr}.
Other gap symmetries are found to be subdominant, giving significantly smaller gap amplitudes. We also limit our considerations to even-frequency gaps since Ref.~\cite{Eliash_Even} did not find solutions with odd-frequency pairing in the linearized equations.

\begin{figure*}[ht]
  \centering
\includegraphics[width=\linewidth]{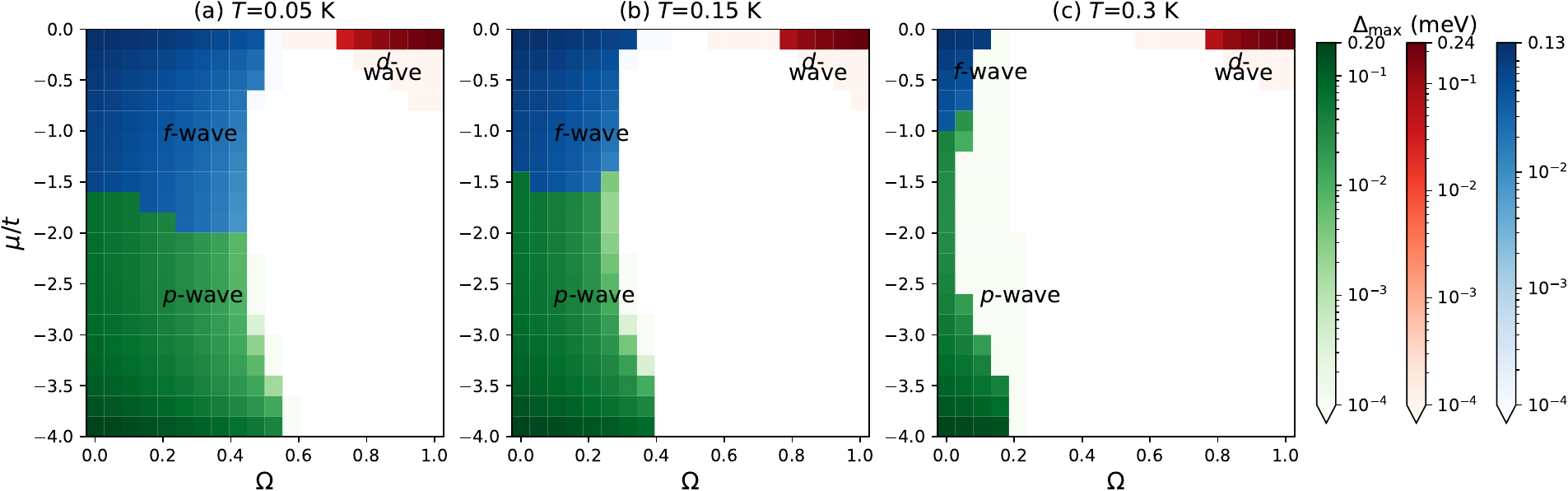}
  \caption{Phase diagram showing the gap symmetry with the largest superconducting gap amplitude $\Delta_{\text{max}}$ in terms of sublattice coupling asymmetry $\Omega$ and chemical potential $\mu$ at different temperatures. The green, blue, and red colors correspond to spin-triplet $p$-wave, spin-triplet $f$-wave, and spin-singlet $d$-wave phases, respectively. The parameters are $t = 1$~eV, $\Bar{J} = 15$~meV, $J_1 = 2$~meV, $J_2 = 0.2J_1$, $K = J_1/10^4$, and $S = 1$.}%
  \label{fig:T_tot}
\end{figure*}

The free energy would provide a way to identify the preferred gap symmetry among the ansatze. 
However, we expect very small differences in the free energy since we consider gaps that are much smaller than the electron bandwidth. Also, the estimate of the free energy will be affected by the FS averages. The superconducting gap provides a condensation energy to the system, which is related to the magnitude of the gap. Choosing functions $\psi(\boldsymbol{k})$ that are normalized with maximum absolute value $1$, the gap amplitude $\Delta_{\text{max}}$ is determined by the zero-imaginary-frequency limit of the gap $\Delta_{s,t}(i\omega_n)$, obtained by solving Eq.~\eqref{eq:2} \cite{Maeland2023EliashCC, Vidberg1977Nov, Marsiglio1988Apr, Aperis2015Aug, Aperis2018Feb, Schrodi2020Mar}. Meanwhile, the gapping of the FS is also affected by the $\boldsymbol{k}$ dependence of the gap. We suggest that $\Delta_{\text{max}}\langle |\psi(\boldsymbol{k})|\rangle_{\text{FS}}$ provides a reasonable indication of the condensation energy, and hence which superconducting symmetry channel that is energetically preferred. When we construct phase diagrams, we find regions of parameter space where gap equations obtained by assuming different ansatze have solutions simultaneously. We then assume that the gap symmetry with the largest value for $\Delta_{\text{max}}\langle |\psi(\boldsymbol{k})|\rangle_{\text{FS}}$ dominates. We compare to the root-mean-square $\Delta_{\text{max}}\sqrt{\langle |\psi(\boldsymbol{k})|^2\rangle_{\text{FS}}}$, which gives the same conclusions. More accurate estimates based on the free energy \cite{Maeland2023EliashCC, Aase2023Dec, Carbotte1990FreeEnergy, Chubukov2020FEspecialized} could be obtained using a full-bandwidth approach to Eliashberg theory and explicitly solving for the momentum dependence of the gap in the entire 1BZ \cite{Aperis2018Feb, Schrodi2020FullBandwidth}. This is a computationally demanding task outside the scope of this paper.

We begin by considering TRS-preserving gap symmetries corresponding to one of the ansatze in Eqs.~\eqref{eq:pwave}-\eqref{eq:fwave}. 
We find nonzero superconducting gap amplitudes at various temperatures in Fig.~\ref{fig:T_tot}, where different colors are utilized to indicate the dominant gap symmetry among $p$-, $d$-, and $f$-wave gaps.
For large sublattice coupling asymmetry (small $\Omega$) and low $\mu$, we find that gaps with spin-triplet $p$-wave symmetry provide the largest gap amplitudes, as indicated by the green region in Fig.~\ref{fig:T_tot}. Within the green region, the largest superconducting gap with spin-triplet $p$-wave symmetry is achieved for $\Omega=0$ and the lowest considered chemical potential $\mu=-3.9t$. 
This is because $\Omega=0$ allows maximal constructive sublattice interference~\cite{Eliash_Even}.  
Furthermore, the small FS at low chemical potential ensures that all scattering processes are of regular type with small magnon scattering momenta. This makes the magnon frequency in the denominator of the magnon propagator small, which explains why the largest gaps are obtained for low chemical potential. 
We find that the $p_x$ and $p_y$ ansatze give degenerate solutions of the Eliashberg equations, as expected from the fourfold rotational symmetry of the square lattice. 
The TRS-broken solutions involving complex gaps will be discussed later. 
We also find solutions of the gap equation in Eq.~\eqref{eq:2} by inserting  an $f$-wave ansatz, but the gap is smaller. Throughout the phase diagram $\langle |\psi_{p_x, p_y}(\boldsymbol{k})|\rangle_{\text{FS}} \approx \langle |\psi_{f_x, f_y}(\boldsymbol{k})|\rangle_{\text{FS}}$, so we simply use the gap amplitude $\Delta_{\text{max}}$ to determine the preferred gap symmetry and distinguish the $p$- and $f$-wave regions of the phase diagrams. 

As $\mu$ increases and approaches half-filling at $\mu = 0$, the FS becomes larger and the emergence of Umklapp scattering changes the situation. We find that degenerate spin-triplet $f_x$- and $f_y$-wave phase dominates since the momentum structure supports subdominant attractive Umklapp scattering processes close to $\boldsymbol{q} = \boldsymbol{Q}$.
These processes are repulsive for $p$-wave pairing. 

The spin flip in the magnon scattering process provides different signs in front of the Eliashberg equations for the spin-singlet and spin-triplet pairing channels ($\zeta_s = -\zeta_t$). 
As a consequence, all scattering processes are repulsive for $s$-wave pairing.
For $\Omega=1$, however, the boosting factor $A^{UU}$ in $D(k-k')$ is maximized for $\boldsymbol{k}' \approx \boldsymbol{k}+\boldsymbol{Q}$, indicating the possibility for superconductivity due to Umklapp scattering. A gap with $d_{x^2-y^2}$-wave symmetry can provide the sign change necessary for generating an attractive interaction to produce superconductivity. This is because $\psi_{d_{x^2-y^2}}(\boldsymbol{k})$ and $\psi_{d_{x^2-y^2}}(\boldsymbol{k}'\approx \boldsymbol{k}+\boldsymbol{Q})$ typically have opposite signs.
This results in a $d$-wave phase for $\mu$ approaching half-filling and large $\Omega$ as shown by the red region. Note that $s$- and $d_{xy}$-wave symmetries cannot provide the sign change necessary for attractive interaction and, therefore, no converging solutions can be found for pure $s$- and $d_{xy}$-wave symmetries.

\begin{figure}[ht]
  \centering
\includegraphics[width=\columnwidth]{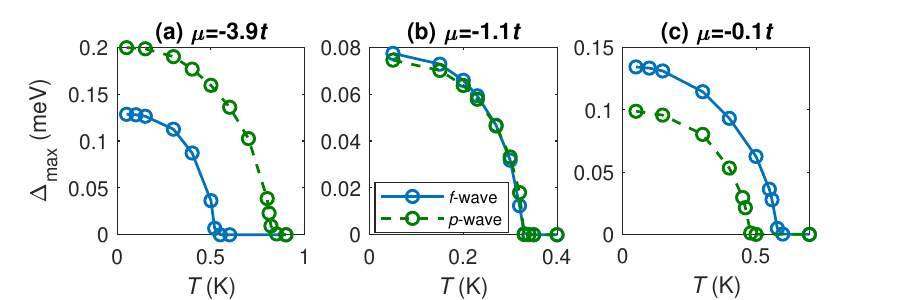}
  \caption{Temperature dependence of the superconducting gap amplitude $\Delta_\text{max}$ at $\Omega=0$ for (a) $\mu=-3.9t$, (b) $\mu=-1.1t$, and (c) $\mu=-0.1t$. The other parameters are the same as in Fig.~\ref{fig:T_tot}.}%
  \label{fig:T_dep}
\end{figure}

Comparing the phase diagrams at different temperatures in Fig.~\ref{fig:T_tot}, it can be seen that the area of each superconducting phase decreases as temperature increases, which corresponds to the general trend that the superconducting gap decreases as temperature increases, and becomes zero at the critical temperature. As a result, a larger boosting factor is required for nonzero superconducting gaps at a given temperature. 
In addition, the chemical potential $\mu$ at the boundary between dominant $p$-wave and $f$-wave pairing increases as temperature increases, 
suggesting that temperature alters the competition between the two types of pairing. 
To further illustrate the temperature dependence of the competition between $p$-wave and $f$-wave pairings, we plot the superconducting gap amplitudes as a function of temperature for different $\mu$ at $\Omega=0$ in Fig.~\ref{fig:T_dep}. At $\mu=-3.9t$, the $p$-wave gap amplitude dominates over the $f$-wave case at all temperatures. In a similar way, the $f$-wave gap dominates at all temperatures at $\mu=-0.1t$. For the intermediate chemical potential $\mu=-1.1t$, a gap amplitude crossing at around $T=0.25$~K is observed, and $f$-wave pairing has the largest gap value below this temperature, while $p$-wave pairing has the largest amplitude above. 
This is consistent with Fig.~\ref{fig:T_tot}, which gives $f$-wave ($p$-wave) pairing for $T=0.15$ K ($T=0.30$ K) at \(\mu = -1.1 t\). 
However, the result has to be interpreted with care. The $f$-wave phase emerges as a consequence of interplay between different types of scattering processes, which may open up for more involved gap structures on the FS than the $p$-wave and $f$-wave ansatze. 
In principle, one could even imagine a crossover between regimes with $p$-wave and $f$-wave behavior, as this can not be described by the simple ansatze.
Importantly, however, the phase diagram clearly captures the increasing importance of the subdominant Umklapp processes when the filling is increased.

\begin{figure}[ht]
  \centering
\includegraphics[width=\columnwidth]{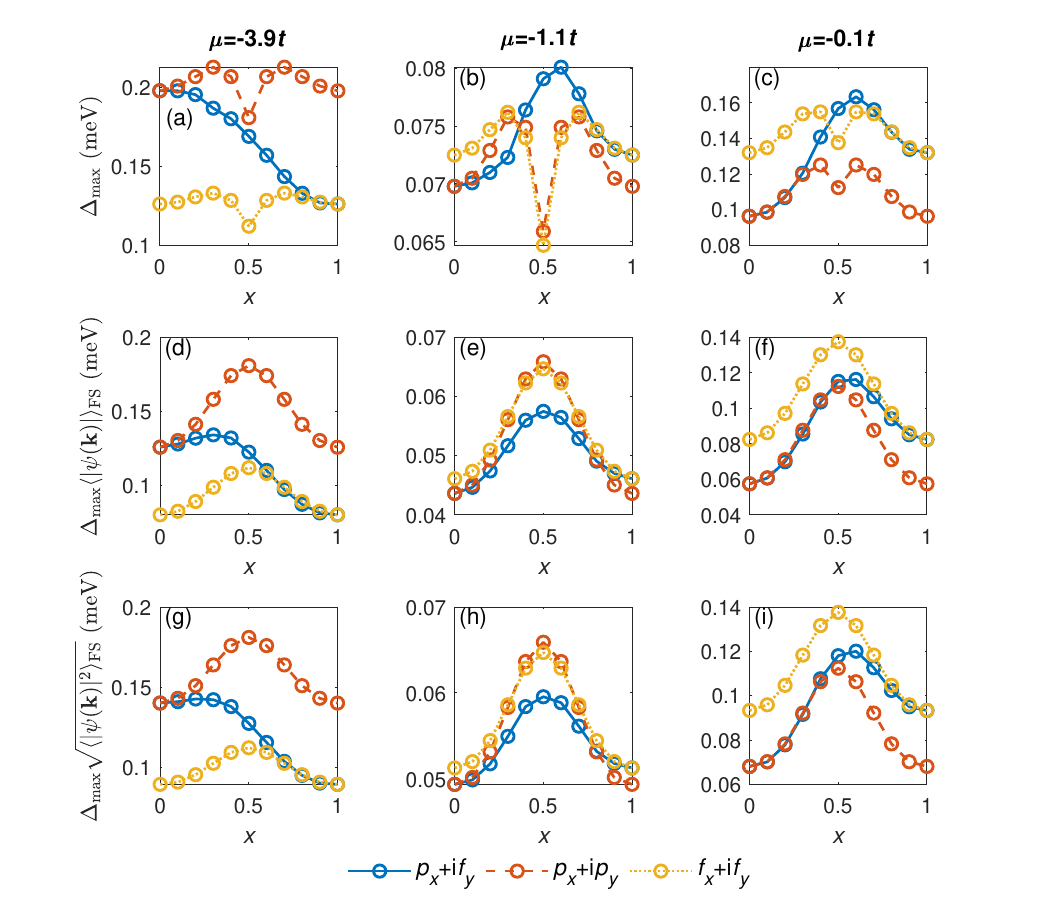}
  \caption{The first row (a-c) shows the superconducting gap amplitude $\Delta_{\text{max}}$ as a function of the mixture ratio $x$ for $\psi(\boldsymbol{k}) = \psi_c(\boldsymbol{k})/\text{max}_{\boldsymbol{k}}|\psi_c(\boldsymbol{k})|$ with $\psi_c(\boldsymbol{k})=(1-x)\psi_{p_x}(\boldsymbol{k}) + ix\psi_{f_y}(\boldsymbol{k})$ for $p_x+if_y$ symmetry, $\psi_c(\boldsymbol{k})=(1-x)\psi_{p_x}(\boldsymbol{k}) + ix\psi_{p_y}(\boldsymbol{k})$ for $p_x+ip_y$ symmetry, and $\psi_c(\boldsymbol{k})=(1-x)\psi_{f_x}(\boldsymbol{k}) + ix\psi_{f_y}(\boldsymbol{k})$ for $f_x+if_y$ symmetry. 
  The second row (d-f) and third row (g-i) plot $\Delta_{\text{max}}\langle|\psi(\boldsymbol{k})|\rangle_{\text{FS}}$ and $\Delta_{\text{max}}\sqrt{\langle|\psi(\boldsymbol{k})|^2\rangle_{\text{FS}}}$ for the same gap symmetries, respectively. 
  In the left column (a,d,g), $\mu=-3.9t$, in the middle column (b,e,h), $\mu=-1.1t$, and in the right column (c,f,i), $\mu=-0.1t$.
  The parameters are $T=0.15 $~K, $\Omega=0$, and otherwise the same as in Fig.~\ref{fig:T_tot}.}%
  \label{fig:mix}
\end{figure}

We now turn to TRS-breaking complex gap functions \cite{Sigrist1991Apr} including the mixtures of different gap symmetries. A free global complex phase choice remains, so we choose $\Delta(i\omega_n)$ to be real and let $\psi(\boldsymbol{k}) = \psi_c(\boldsymbol{k})/\text{max}_{\boldsymbol{k}}|\psi_c(\boldsymbol{k})|$ handle both the momentum dependence and the complex phase. We consider chiral $p$-wave [i.e., $p_x+ip_y$, $\psi_c(\boldsymbol{k})=(1-x)\psi_{p_x}(\boldsymbol{k}) + ix\psi_{p_y}(\boldsymbol{k})$], chiral $f$-wave [i.e., $f_x+if_y$, $\psi_c(\boldsymbol{k})=(1-x)\psi_{f_x}(\boldsymbol{k}) + ix\psi_{f_y}(\boldsymbol{k})$], and $p_x+if_y$-wave symmetry [$\psi_c(\boldsymbol{k})=(1-x)\psi_{p_x}(\boldsymbol{k}) + ix\psi_{f_y}(\boldsymbol{k})$]. These symmetries give nodeless gaps. The first row in Fig.~\ref{fig:mix} shows the gap amplitude as a function of the mixture ratio $x$ at $\Omega=0$ for different $\mu$. For chiral $p$-wave and chiral $f$-wave, the gap amplitude is symmetric with respect to $x=0.5$, due to the degeneracy between $p_x$ $(f_x)$ and $p_y$ $(f_y)$ solutions. The dips at $x=0.5$ are introduced by rather sharp jumps up to 1 for $\langle |\psi(\boldsymbol{k})|^2 \rangle_{\text{FS}}$ appearing in the denominator of Eq.~\eqref{eq:2}. $\langle |\psi(\boldsymbol{k})|^2 \rangle_{\text{FS}} \approx 0.5$ for $x=0$ and $x=1$, $\langle |\psi(\boldsymbol{k})|^2 \rangle_{\text{FS}} \approx 0.7$ for $x=0.4$ and $x=0.6$, and $\langle |\psi(\boldsymbol{k})|^2 \rangle_{\text{FS}} =1$ for $x=0.5$. At the same time, a large $\langle |\psi(\boldsymbol{k})|^2 \rangle_{\text{FS}}$ should provide a better gapping of the FS, so the gap amplitude alone is not the best indication of which state is preferred.

The second row in Fig.~\ref{fig:mix} shows $\Delta_{\text{max}}\langle |\psi(\boldsymbol{k})|\rangle_{\text{FS}}$ for the same gap symmetries. For each considered symmetry, this quantity is optimized at certain mixtures $0<x<1$. This result indicates that the TRS-breaking gaps are energetically preferred over the TRS-preserving gaps. At $\mu = -3.9t$ we predict that $p_x+ip_y$ with $x \approx 0.5$ gives the greatest condensation energy among the considered gap symmetries. For $\mu = -1.1t$ there is a close competition between chiral $p$-wave and chiral $f$-wave at $x=0.5$ with chiral $p$-wave giving a slightly greater $\Delta_{\text{max}}\langle |\psi(\boldsymbol{k})|\rangle_{\text{FS}}$. Interestingly, at $\mu=-1.1t$, $f$-wave seems preferred over $p$-wave when restricting to TRS-preserving gaps. At $\mu = -0.1t$ the results in Fig.~\ref{fig:mix} suggest that chiral $f$-wave with mixture ratio $x \approx 0.5$ is energetically preferred. On the other hand, $p_x+if_y$ appears to be a subdominant candidate for the superconducting gap symmetry at all considered parameters. Even though this gap symmetry produces the largest gap amplitude at $\mu = -1.1t$ and $\mu = -0.1t$ in Figs.~\ref{fig:mix}(b) and (c), it provides a less efficient gapping of the FS compared to chiral $p$- and $f$-wave. The last row in Fig.~\ref{fig:mix} shows $\Delta_{\text{max}}\sqrt{\langle |\psi(\boldsymbol{k})|^2\rangle_{\text{FS}}}$, which only has slight quantitative changes from $\Delta_{\text{max}}\langle |\psi(\boldsymbol{k})|\rangle_{\text{FS}}$, and gives the same indication for the preferred symmetries.

\begin{figure}[ht]
  \centering
\includegraphics[width=\columnwidth]{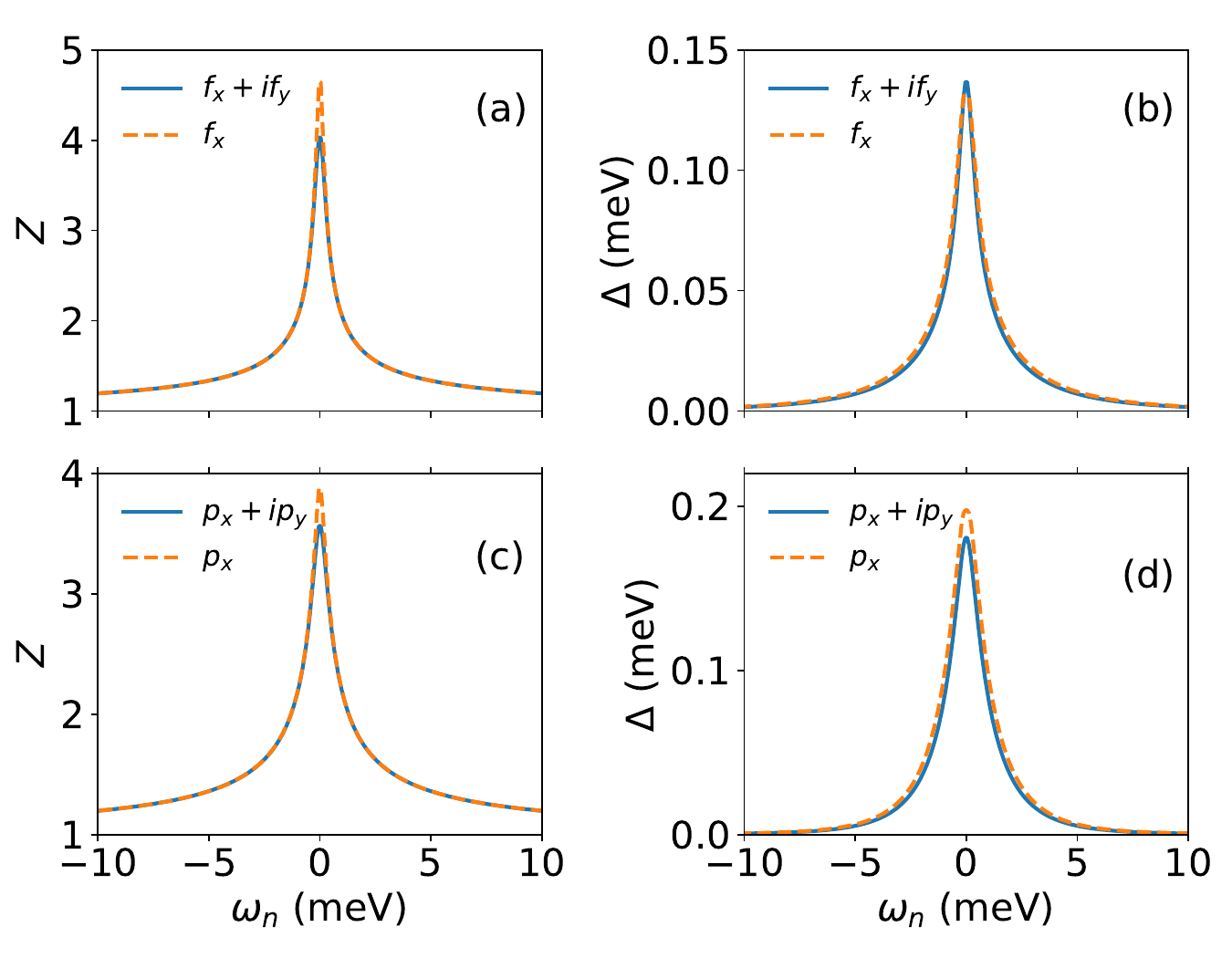}
  \caption{The frequency dependence of (a) the electron renormalization $Z(i\omega_n)$ and (b) the gap $\Delta(i\omega_n)$ at $\mu = -0.1t$ with $\psi(\boldsymbol{k})=\psi_{f_x}(\boldsymbol{k})$ and $\psi(\boldsymbol{k})=\psi_{f_x}(\boldsymbol{k}) + i\psi_{f_y}(\boldsymbol{k})$ (This is equivalent to $x=0$ and $x=0.5$ for $f_x+if_y$ symmetry described in Fig.~\ref{fig:mix}). (c) $Z(i\omega_n)$ and (d) $\Delta(i\omega_n)$ at $\mu = -3.9t$ with $\psi(\boldsymbol{k})=\psi_{p_x}(\boldsymbol{k})$ and $\psi(\boldsymbol{k})=\psi_{p_x}(\boldsymbol{k}) + i\psi_{p_y}(\boldsymbol{k})$ (This is equivalent to $x=0$ and $x=0.5$ for $p_x+ip_y$ symmetry described in Fig.~\ref{fig:mix}). The parameters are $T=0.15 $~K, $\Omega=0$ and otherwise the same as in Fig.~\ref{fig:T_tot}.}%
  \label{fig:ZDelta}
\end{figure}

Figures \ref{fig:ZDelta}(a) and \ref{fig:ZDelta}(b) plot the frequency dependence of the electron renormalization function $Z(i\omega_n)$ and the gap $\Delta(i\omega_n)$ for $f_x$- and chiral $f$-wave gaps at $\mu = -0.1t$. Note that $Z(i\omega_n) = 1$ in the case of zero renormalization, and therefore Fig.~\ref{fig:ZDelta} indicates the presence of significant many-body effects. Since $Z(i\omega_n)$ enters in the denominator in Eq.~\eqref{eq:2}, $Z(i\omega_n)>1$ results in a prediction of smaller gap amplitudes than in BCS theory \cite{Sun2023Aug}. Furthermore, the better gapping of the FS for chiral $f$-wave gap results in lower $Z(i\omega_n)$ compared to the $f_x$-wave gap case. The reduction is due to the fact that the gap enters in the denominator in the equation for $Z(i\omega_n)$. A reduction in $Z(i\omega_n)$ allows for a larger gap amplitude for chiral $f$-wave than $f_x$-wave, which does not seem possible without taking renormalization into account. 
Figures \ref{fig:ZDelta}(c) and \ref{fig:ZDelta}(d) plot $Z(i\omega_n)$ and $\Delta(i\omega_n)$ for $p_x$- and chiral $p$-wave gaps at $\mu = -3.9t$. In that case, the reduction in $Z(i\omega_n)$ is not large enough to allow larger gap amplitude for $p_x+ip_y$-wave gap than $p_x$-wave gap. Note from Figs.~\ref{fig:mix}(a)-(c) that the typical case is that the chiral gaps with $x = 0.5$ have lower gap amplitude than the TRS preserving gaps. Still, the gap amplitude is not reduced by a very large factor, and the better gapping of the FS provided by the chiral gaps, results in a clear indication that they are energetically preferred in Figs.~\ref{fig:mix}(d)-(i). In Appendix \ref{app:alternate} we provide an additional explanation for why the gap amplitude is not reduced by as much as one might naively expect when going from $p_x$- to $p_x+ip_y$-wave gap symmetries. This additional explanation based on the momentum dependence works in conjunction with the explanation related to the renormalization $Z(i\omega_n)$.

The exotic possibility of chiral $d$-wave pairing has attracted significant attention~\cite{Senthil1999dwave, Vojta2000dwave, Black-Schaffer2012dwave, Black-Schaffer2014dwave, Fischer2014dwave, Holmvall2023dwave}. In our case, however, the repulsiveness of pure $d_{xy}$-wave pairing prevents this possibility, as we discuss in more detail in Appendix \ref{app:alternate}.

When $0<x<1$ all the gaps considered in Fig.~\ref{fig:mix} break TRS since there is a phase difference between the real and imaginary parts \cite{Sigrist1991Apr}. They also result in spinful topological superconductivity. This is well known for chiral $p$-, $d$-, and $f$-wave superconductors \cite{Schnyder2008tenfold, Huang2014chiralpdf, Black-Schaffer2014dwave}. We also study the possibility of $p_x+if_y$-wave symmetry. This is a topological superconductor since a $p_x-ip_y$-wave gap can be continuously deformed into a $p_x+if_y$-wave gap without closing the bulk gap in the superconductor. For completeness, we calculate the bulk topological invariant, discuss the symmetry classification, and prove the existence of edge states for a $p_x+if_y$-wave gap within a mean-field BCS formalism in Appendix \ref{app:TSC}. 

Spin-orbit coupling (SOC) is often an integral part of potential platforms for topological superconductivity \cite{Kitaev2001Oct, Oreg2010Oct, MicrosoftTGP, Hell2017JJTSC, Pientka2017JJTSC, Lesser2022JJTSC, TopoSCandSkRev, Maeland2023Apr, Maeland2023Dec, Bostrom2023Dec}.
Beyond a very weak easy-axis anisotropy in the AFMI, which hardly plays any role in inducing the topologically nontrivial gaps, SOC does not enter in our model. Also, the effective coupling in the Eliashberg equations, $V^2 D(k-k')$, is real. The possibility of complex superconducting gaps is therefore not imposed by the interaction itself, rather, the superconducting state may spontaneously break TRS to lower its energy. A fourfold-symmetric lattice, such that $p_x$ ($f_x$) and $p_y$ ($f_y$) are degenerate facilitates this situation. However, we can speculate that this situation is not limited to lattices where  $p_x$ ($f_x$) and $p_y$ ($f_y$) are exactly degenerate. As long as both are possible solutions of the gap equations, maybe with slightly different $T_c$, TRS-breaking linear combinations may be preferred at low temperatures since they provide a better gapping of the FS.

\section{Conclusion}
In this paper, we used a strong-coupling approach to study superconductivity mediated by antiferromagnetic magnons in an AFMI/NM/AFMI trilayer. We derived Fermi-surface-averaged nonlinear Eliashberg equations, which can describe superconductivity at any temperature.
At low temperatures below the critical temperature, we found states with different superconducting gap symmetries ($p$-, $f$-, and $d$-wave) in the $(\mu,\Omega)$-phase diagram, in agreement with the phase diagrams achieved at the critical temperature and at zero temperature. 
Comparisons of the phase diagrams at different temperatures and the temperature dependence of the gaps at different $\mu$  revealed that temperature affects the competition between $p$-wave and $f$-wave pairing, and that $p$-wave pairing is more likely to dominate when the temperature increases from zero. 
In addition to time-reversal symmetric superconducting gap functions, we also investigated time-reversal-symmetry-breaking complex gap functions. The complex gap functions involved a mixture of different superconducting gap symmetries. We considered a Fermi surface average of the gap amplitude as an indication of the condensation energy. From this, we found that time-reversal-symmetry breaking, spinful topological superconductivity is possible.  

\begin{acknowledgments} 
We acknowledge funding from the Research Council of Norway (RCN) through its Centres of Excellence funding scheme Project No.~262633, ``QuSpin," and RCN Project No.~323766, ``Equilibrium and out-of-equilibrium quantum phenomena in superconducting hybrids with antiferromagnets and topological insulators." 
This work was partially supported by the Swiss National Science Foundation.
\end{acknowledgments}

\appendix

\section{Alternate approach to complex gaps} \label{app:alternate}
In this appendix we explore an alternate approach to Fermi surface averaged Eliashberg equations in the case of complex gaps than the one employed in the main text.
The Eliashberg equation for $\phi_{s,t}(k) = \phi_{s,t}^R(k) + i \phi_{s,t}^I(k)$ could also have been written as
\begin{align}
    \phi_{s,t}^R(k) = -\zeta_{s,t}V^2\frac{1}{\beta}\sum_{k'}D(k-k')\frac{\phi_{s,t}^R(k')}{\Theta(k')}, \\
    \phi_{s,t}^I(k) = -\zeta_{s,t}V^2\frac{1}{\beta}\sum_{k'}D(k-k')\frac{\phi_{s,t}^I(k')}{\Theta(k')}.
\end{align}
We can apply FS averages to these equations by assuming $\phi_{s,t}^R(k) = \phi_{s,t}^R(i\omega_n)\psi^R(\boldsymbol{k})$ and $\phi_{s,t}^I(k) = \phi_{s,t}^I(i\omega_n)\psi^I(\boldsymbol{k})$. The gap function is $\Delta_{s,t}(k) = \Delta_{s,t}^R(k)+i\Delta_{s,t}^I(k) = \Delta_{s,t}^R(i\omega_n)\psi^R(\boldsymbol{k}) + i \Delta_{s,t}^I(i\omega_n)\psi^I(\boldsymbol{k})$ and we can choose real functions $\psi^R(\boldsymbol{k})$ and $\psi^I(\boldsymbol{k})$ to explore complex gaps. 
The FS-averaged equations are then
\begin{align}
   &Z(i\omega_n)=1-\frac{V^2\pi}{\beta N_F \omega_n}\sum_{\boldsymbol{k},\boldsymbol{k}{'},\omega_{n{'}}}\delta(\xi_{\boldsymbol{k}})\delta(\xi_{\boldsymbol{k}{'}})D(k-k{'})\notag\\
   &\times\frac{\omega_{n{'}}}{\sqrt{\omega_{n{'}}^2+|\Delta_{s,t}^R(i\omega_{n'})\psi^R(\boldsymbol{k}') + i \Delta_{s,t}^I(i\omega_{n'})\psi^I(\boldsymbol{k}')|^2}}, \label{eq:EliFSalt1}
\end{align}
\begin{align}
   &\Delta_{s,t}^R(i\omega_n)=\frac{-\zeta_{s,t}V^2}{\langle [\psi^R(\boldsymbol{k})]^2 \rangle_{\text{FS}} Z(i\omega_n)}\frac{\pi}{\beta N_F}\sum_{\boldsymbol{k},\boldsymbol{k}{'},\omega_{n{'}}}\delta(\xi_{\boldsymbol{k}})\delta(\xi_{\boldsymbol{k}{'}})\notag\\
   &\times\frac{\psi^R(\boldsymbol{k})D(k-k{'})\psi^R(\boldsymbol{k}{'})\Delta_{s,t}^R(i\omega_{n{'}})}{\sqrt{\omega_{n{'}}^2+|\Delta_{s,t}^R(i\omega_{n'})\psi^R(\boldsymbol{k}') + i \Delta_{s,t}^I(i\omega_{n'})\psi^I(\boldsymbol{k}')|^2}}, \label{eq:EliFSalt2}
\end{align}
\begin{align}
   &\Delta_{s,t}^I(i\omega_n)=\frac{-\zeta_{s,t}V^2}{\langle [\psi^I(\boldsymbol{k})]^2 \rangle_{\text{FS}} Z(i\omega_n)}\frac{\pi}{\beta N_F}\sum_{\boldsymbol{k},\boldsymbol{k}{'},\omega_{n{'}}}\delta(\xi_{\boldsymbol{k}})\delta(\xi_{\boldsymbol{k}{'}})\notag\\
   &\times\frac{\psi^I(\boldsymbol{k})D(k-k{'})\psi^I(\boldsymbol{k}{'})\Delta_{s,t}^I(i\omega_{n{'}})}{\sqrt{\omega_{n{'}}^2+|\Delta_{s,t}^R(i\omega_{n'})\psi^R(\boldsymbol{k}') + i \Delta_{s,t}^I(i\omega_{n'})\psi^I(\boldsymbol{k}')|^2}}. \label{eq:EliFSalt3}
\end{align}
With this equation set we are not able to control the mixture ratio $x$ in chiral $p$-wave, chiral $f$-wave, and $p_x+if_y$ gaps, where $\psi^R(\boldsymbol{k}) \neq \psi^I(\boldsymbol{k})$. This is because the frequency dependence of the real and imaginary parts of the gaps are solved in separate, coupled equations. For TRS-preserving gaps, where $\psi^R(\boldsymbol{k}) = \psi^I(\boldsymbol{k})$, the two equations for $\Delta_{s,t}^R(i\omega_n)$ and $\Delta_{s,t}^I(i\omega_n)$ simply reflect the free complex phase choice of the superconducting gap, and one can choose the solution where $\Delta_{s,t}^I(i\omega_n) = 0$. This gives real gaps, as we chose in the main text for the TRS-preserving gaps.

Equation \eqref{eq:EliFSalt2} provides an understanding of why adding an imaginary part to the gap does not necessarily lead to a strong reduction in the gap amplitude, as discussed in relation to Figs.~\ref{fig:mix} and \ref{fig:ZDelta}. Imagine we have a solution of $\Delta_t^R(i\omega_n)$ where the gap is real with $\psi^R(\boldsymbol{k}) = \psi_{p_x}(\boldsymbol{k})$. Let us now try to introduce an imaginary part $\Delta_t^I(i\omega_n)$ and set $\psi^I(\boldsymbol{k}) = \psi_{p_y}(\boldsymbol{k})$, giving a $p_x+ip_y$-wave gap. Then, $|\Delta_{t}^R(i\omega_{n'})\psi^R(\boldsymbol{k}') + i \Delta_{t}^I(i\omega_{n'})\psi^I(\boldsymbol{k}')|^2$ in the denominator increases, so that a smaller gap amplitude of $\Delta_t^R(i\omega_n)$ should be expected. However, $\psi^R(\boldsymbol{k}')$ enters in the numerator. When $\psi^I(\boldsymbol{k}')$ is large, $\psi^R(\boldsymbol{k}')$ is small, and vice versa. Therefore, the effect on the gap amplitude of adding an imaginary part with a different phase is not as significant as adding an imaginary part with the same phase. Also, remember that $Z(i\omega_n)$ is affected by the gap symmetry, which in turn affects the gap amplitude.

If we set $\psi^R(\boldsymbol{k}) = \psi_{p_x}(\boldsymbol{k})$ and $\psi^I(\boldsymbol{k}) = \psi_{p_y}(\boldsymbol{k})$ and start our iterations from $\Delta_{t}^R(i\omega_n) \neq \Delta_{t}^I(i\omega_n)$ the solutions converge toward $\Delta_{t}^R(i\omega_n) = \Delta_{s,t}^I(i\omega_n)$. This corresponds to chiral $p$-wave with mixture ratio $x=0.5$. The gap amplitudes $\Delta^R_{\text{max}} = \Delta^I_{\text{max}}$ are the same as the result in the main text, as is the conclusion that equal mixing should be preferred. Also, $\langle|\Delta^R_{\text{max}}\psi^R(\boldsymbol{k}') + i \Delta^I_{\text{max}}\psi^I(\boldsymbol{k}')|\rangle_{\text{FS}}$ equals to $\Delta_{\text{max}}\langle |\psi(\boldsymbol{k})|\rangle_{\text{FS}}$ in Fig.~\ref{fig:mix} for $x = 0.5$. Similar considerations apply to chiral $f$-wave, while for $p_x+if_y$-wave gaps, solutions with a $\mu$-dependent mixture ratio $x$ are found with the alternate equations. The method in the main text allows us to consider a wider range of possible gap symmetries, although it appears some of them are not necessarily solutions of the original Eliashberg equations in Eqs.~\eqref{Eli0_1} and \eqref{Eli0_2}. This must be kept in mind when interpreting the results. One can view the mixture ratio $x$ in Fig.~\ref{fig:mix} as a variational parameter we use to search for an optimal solution for a class of gap symmetries.

\begin{figure}[ht]
  \centering
\includegraphics[width=0.6\columnwidth]{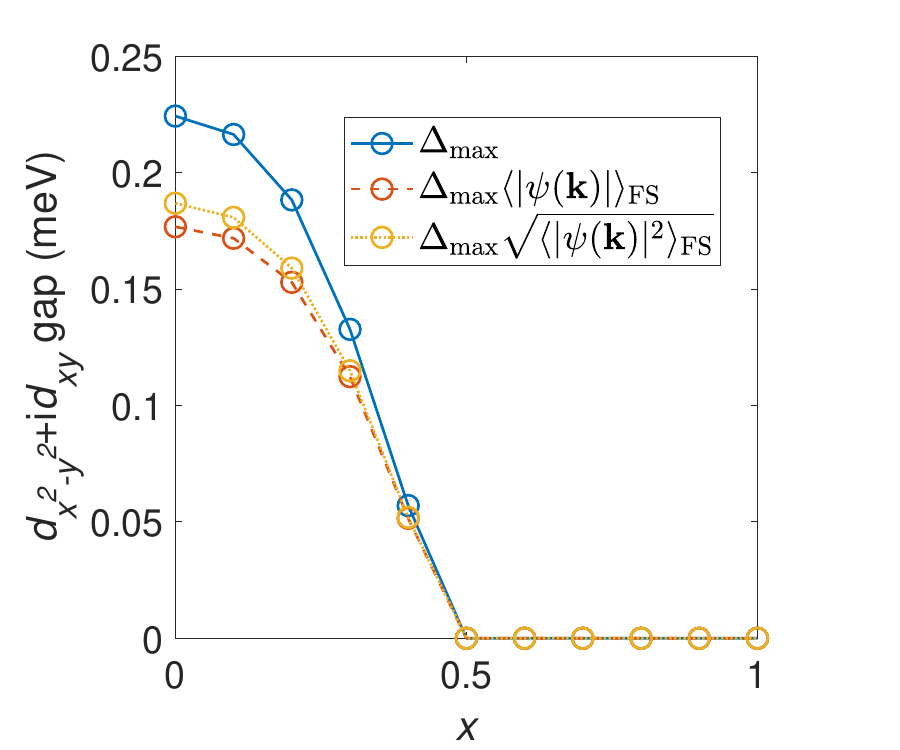}
  \caption{Superconducting gap amplitude as a function of the mixture ratio $x$ for $\psi(\boldsymbol{k}) = \psi_c(\boldsymbol{k})/\text{max}_{\boldsymbol{k}}|\psi_c(\boldsymbol{k})|$ with $\psi_c(\boldsymbol{k})=(1-x)\psi_{d_{x^2-y^2}}(\boldsymbol{k}) + ix\psi_{d_{xy}}(\boldsymbol{k})$ at $\mu=-0.1t$, $\Omega=1$ and $T=0.15 $~K. The other parameters are the same as in Fig.~\ref{fig:T_tot}.}%
  \label{fig:dmix}
\end{figure}

With the equations in the main text, we do in fact get converging solutions to the Eliashberg equations with chiral $d$-wave gaps.
Figure \ref{fig:dmix} shows the gap amplitude as a function of the mixture ratio $x$ for $d_{x^2-y^2}+id_{xy}$-wave gaps with $\psi_c(\boldsymbol{k})=(1-x)\psi_{d_{x^2-y^2}}(\boldsymbol{k}) + ix\psi_{d_{xy}}(\boldsymbol{k})$. As long as $d_{x^2-y^2}$-wave dominates, i.e., $x<0.5$, converging solutions are found to the Eliashberg equations. The gap amplitude decreases away from $x=0$, indicating that pure $d_{x^2-y^2}$-wave is preferred. If we instead consider the Eliashberg equations on the form in Eqs.~\eqref{eq:EliFSalt1}-\eqref{eq:EliFSalt3} with $\psi^R(\boldsymbol{k}) = \psi_{d_{x^2-y^2}}(\boldsymbol{k})$ and $\psi^I(\boldsymbol{k}) = \psi_{d_{xy}}(\boldsymbol{k})$ we find that $\Delta_{s}^I(i\omega_n) = 0$ is the only possible solution. Due to the momentum structure of the interaction discussed in the main text regarding pure $d_{xy}$-wave gaps, $\Delta_{s}^I(i\omega_n)$ changes sign at each step in the iteration, preventing nonzero solutions. Again, this not only indicates that pure $d_{x^2-y^2}$-wave gap is preferred over chiral $d$-wave, but it also indicates that chiral $d$-wave is not a possible solution of the original Eliashberg equations in Eqs.~\eqref{Eli0_1} and \eqref{Eli0_2}.

\section{Topological superconductivity} \label{app:TSC}
\subsection{Bulk topological invariant}
To explore the topology of the superconducting gaps studied in the main text, we employ a mean-field BCS description \cite{Sigrist1991Apr} of the induced superconducting state in the NM.
We consider two kinds of gaps: spin singlet $\Delta_{\uparrow\downarrow}^{O(s)} = (\Delta_{\uparrow\downarrow}-\Delta_{\downarrow\uparrow})/2$ and spin triplet $\Delta_{\uparrow\downarrow}^{E(s)} = (\Delta_{\uparrow\downarrow}+\Delta_{\downarrow\uparrow})/2$ \cite{Sigrist1991Apr, Maeland2023Apr}. We do not consider their coexistence, so in the first case $\Delta_{\downarrow\uparrow} = -\Delta_{\uparrow\downarrow}$ while in the second case $\Delta_{\downarrow\uparrow} = \Delta_{\uparrow\downarrow}$. The Bogoliubov-de Gennes (BdG) Hamiltonian for the bulk can be written as
\begin{equation}
    H = \frac{1}{2}\sum_{\boldsymbol{k}} \boldsymbol{c}_{\boldsymbol{k}}^\dagger H_{\boldsymbol{k}} \boldsymbol{c}_{\boldsymbol{k}} \label{eq:BdG}
\end{equation}
with $\boldsymbol{c}_{\boldsymbol{k}}^\dagger = (c_{\boldsymbol{k}\uparrow}^\dagger, c_{\boldsymbol{k}\downarrow}^\dagger, c_{-\boldsymbol{k}\uparrow}, c_{-\boldsymbol{k}\downarrow})$ and
\begin{equation}
    H_{\boldsymbol{k}}=\begin{pmatrix}
        \epsilon_{\boldsymbol{k}} & 0 & 0 & \Delta_{\boldsymbol{k}} \\
        0 & \epsilon_{\boldsymbol{k}} & \zeta_{s,t} \Delta_{\boldsymbol{k}} & 0 \\
        0 & \zeta_{s,t}\Delta_{\boldsymbol{k}}^* & -\epsilon_{\boldsymbol{k}} & 0 \\
        \Delta_{\boldsymbol{k}}^* & 0 & 0 & -\epsilon_{\boldsymbol{k}}
    \end{pmatrix}.
\end{equation}
This can be rewritten as two $2\times 2$ sectors
\begin{align}
    H &= \frac{1}{2}\sum_{\boldsymbol{k}} (c_{\boldsymbol{k}\uparrow}^\dagger, c_{-\boldsymbol{k}\downarrow}) \begin{pmatrix}
        \epsilon_{\boldsymbol{k}} & \Delta_{\boldsymbol{k}} \\
        \Delta_{\boldsymbol{k}}^* & -\epsilon_{\boldsymbol{k}}
    \end{pmatrix} \begin{pmatrix}
        c_{\boldsymbol{k}\uparrow} \\
        c_{-\boldsymbol{k}\downarrow}^\dagger
    \end{pmatrix}\notag\\&+ \frac{1}{2}\sum_{\boldsymbol{k}} (c_{\boldsymbol{k}\downarrow}^\dagger, c_{-\boldsymbol{k}\uparrow}) \begin{pmatrix}
        \epsilon_{\boldsymbol{k}} & \zeta_{s,t}\Delta_{\boldsymbol{k}} \\
        \zeta_{s,t}\Delta_{\boldsymbol{k}}^* & -\epsilon_{\boldsymbol{k}}
    \end{pmatrix} \begin{pmatrix}
        c_{\boldsymbol{k}\downarrow} \\
        c_{-\boldsymbol{k}\uparrow}^\dagger
    \end{pmatrix}. \label{eq:2x2sectors}
\end{align}
Each sector now has a matrix form that can be written $h_{\boldsymbol{k}} = \boldsymbol{d}_{\boldsymbol{k}} \cdot \boldsymbol{\sigma}$. Then, the Chern number is \cite{Hasan2010BulkTop}
\begin{equation}
    C = \frac{1}{4\pi} \int d\boldsymbol{k} \hat{d}_{\boldsymbol{k}} \cdot (\partial_{k_x} \hat{d}_{\boldsymbol{k}} \times \partial_{k_y} \hat{d}_{\boldsymbol{k}}),
\end{equation}
i.e., the winding number of $\hat{d}_{\boldsymbol{k}} = \boldsymbol{d}_{\boldsymbol{k}}/|\boldsymbol{d}_{\boldsymbol{k}}|$. The integral is over the 1BZ, for which we still consider a square lattice. In the first sector, $\boldsymbol{d}_{\boldsymbol{k}} = (\Re \Delta_{\boldsymbol{k}}, -\Im \Delta_{\boldsymbol{k}}, \epsilon_{\boldsymbol{k}})$. In the other sector, $\boldsymbol{d}_{\boldsymbol{k}} = (\zeta_{s,t}\Re \Delta_{\boldsymbol{k}}, -\zeta_{s,t}\Im \Delta_{\boldsymbol{k}}, \epsilon_{\boldsymbol{k}})$. Changing the sign of two entries does not change the winding, so both sectors give the same Chern number. The total Chern number is the sum, i.e., $C_{\text{tot}} = 2C$. Alternatively we could consider the total Chern number of the two degenerate bands below the FS for the $4\times 4$ BdG Hamiltonian following Refs.~\cite{Sato2014BulkTopSpinful, Mochol-Grzelak2018BulkTopDegen}. Note that this kind of topological superconductor with broken TRS and two bands below the FS is called a spinful topological superconductor \cite{Schnyder2008tenfold, Sato2014BulkTopSpinful}. We checked that both approaches give the same result, as expected \cite{sachdev2023quantum}.

In the case of spin-triplet gaps, $p_x+ip_y$-wave gives $C = 1$ and $C_{\text{tot}} = 2$, $f_x + if_y$-wave gives $C = 3$ and $C_{\text{tot}} = 6$, and $p_x + if_y$-wave gives $C = -1$ and $C_{\text{tot}} = -2$. $p_x + if_y$-wave is in the same topological phase as $p_x-ip_y$ since they can be continuously deformed into each other without closing the bulk gap. Since we carefully designed our trilayer to give zero spin splitting of the electron bands, $\epsilon_{\boldsymbol{k},\uparrow} = \epsilon_{\boldsymbol{k},\downarrow}$, the $2\times2$ sectors in Eq.~\eqref{eq:2x2sectors} retain particle-hole symmetry. As a result, they fall in the symmetry class D of topological insulators and superconductors  \cite{Altland1997tenfold, Schnyder2008tenfold}. Hence, one would expect Majorana bound states on edges and in the core of vortices. However, since the total Chern number is even, the Majoranas come in pairs. These pairs of Majoranas can be recombined into Dirac fermions \cite{Sato2014BulkTopSpinful}. A large part of the allure of topological superconductors is the promise of Majorana bound states whose non-Abelian exchange statistics may facilitate robust quantum computation \cite{Bernevig2013, TopoSCrevSato, Leijnse2012TSCrev, TopoQuantumCompRevModPhys}. Ways to retain the Majorana character in spinful topological superconductors, e.g., by imposing additional mirror symmetries, have been discussed \cite{Sato2014BulkTopSpinful}.

For spin-singlet gaps, $d_{x^2-y^2}+id_{xy}$-wave gives $C = 2$ and $C_{\text{tot}} = 4$. No spin splitting of the electrons along with the spin-singlet symmetry lands this type of topological superconductor in symmetry class C \cite{Schnyder2008tenfold, Black-Schaffer2014dwave}. Again, zero-energy edge states are expected, and ways to generate Majorana bound states at edges of chiral $d$-wave superconductors through additional modifications have been suggested \cite{Black-Schaffer2012dwave, Black-Schaffer2014dwave}. The possibility of Majorana bound states in the core of vortices depends on the topology in one dimension lower than the system itself, since it is a point defect \cite{TeoKane2010defectdm1}. Class C has no topologically nontrivial states in 1D such that Majoranas are not expected in the core of vortices \cite{Black-Schaffer2014dwave}. However, heterostructures involving chiral $d$-wave superconductors and materials with significant SOC may host unpaired Majorana bound states in the core of vortices \cite{Mercado2022Majoranadwave, Margalit2022Majoranadwave}.

\begin{figure}[tbh]
  \centering
  \includegraphics[width=\linewidth]{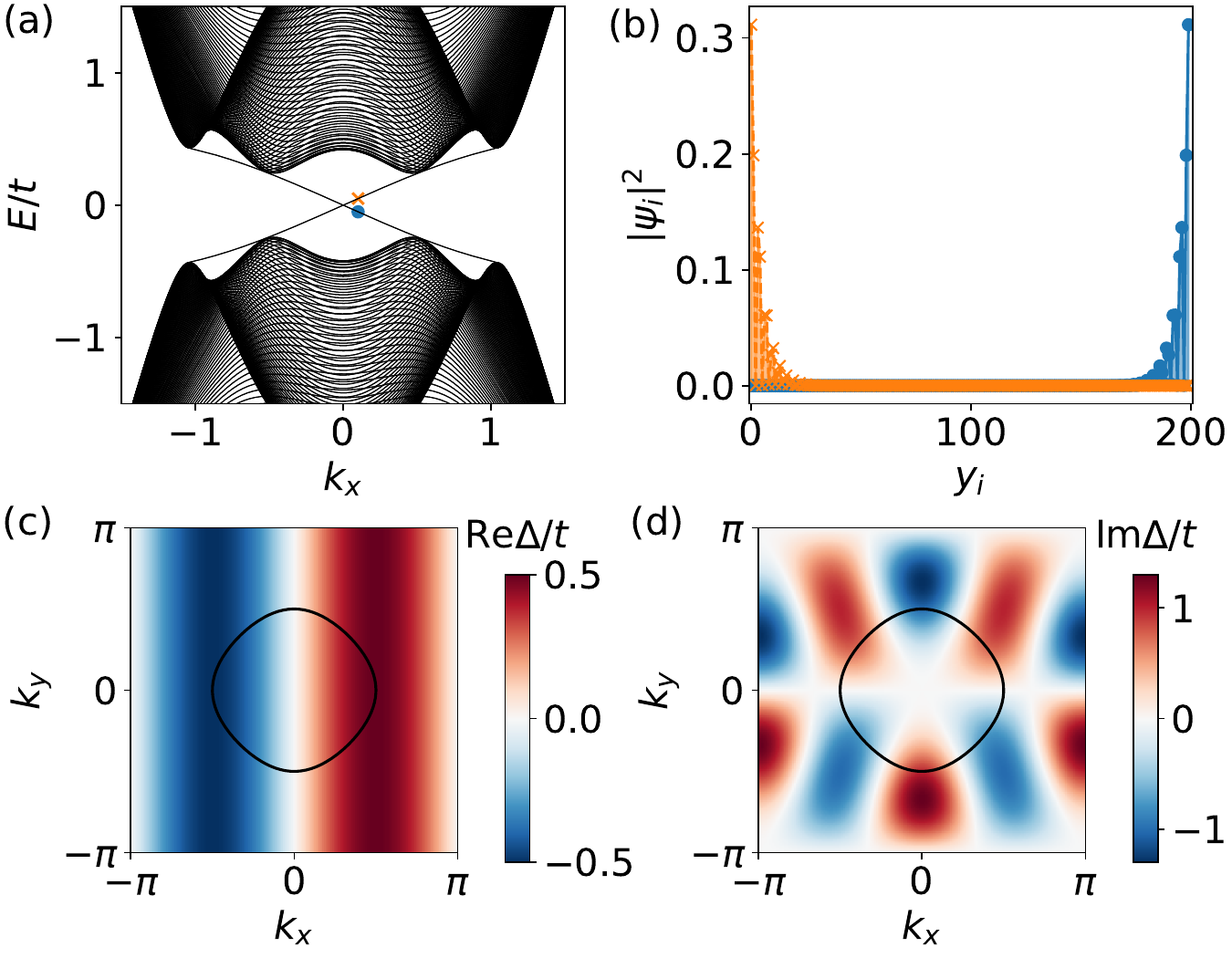}
  \caption{(a) Electron bands along $k_x$ in a ribbon geometry. Two of the states crossing the bulk gap are shown with markers. (b) Weight of the two states marked in (a) along $y_i$. These states are localized at separate edges. (c) Real and (d) imaginary part of $\Delta_{\boldsymbol{k}}$ in Eq.~\eqref{eq:Deltakrealspace}. The black line shows the FS at the chosen chemical potential showing that the considered gap is $p_x+if_y$-wave. The parameters are $\mu = -2t$, $\Delta_p = \Delta_f = 0.5t$, and $N_y = 200$.}%
  \label{fig:edge}
\end{figure}

\subsection{Edge states}
Consider the real-space Hamiltonian
\begin{widetext}
\begingroup
\allowdisplaybreaks
\begin{align}
    H =& -t\sum_{\langle i,j \rangle} c_{i\sigma}^\dagger c_{j\sigma} - \mu \sum_{i\sigma} c_{i\sigma}^\dagger c_{i\sigma} + \frac{\Delta_p}{4i} \sum_{i\sigma} [c_{i\sigma}^\dagger c_{i+\hat{x},-\sigma}^\dagger - c_{i\sigma}^\dagger c_{i-\hat{x}, -\sigma}^\dagger + c_{i\sigma} c_{i+\hat{x},-\sigma} - c_{i\sigma} c_{i-\hat{x}, -\sigma} ] \nonumber \\
    &+ \frac{\Delta_f }{8} \sum_{i\sigma} [c_{i\sigma}^\dagger c_{i+\hat{x}+2\hat{y},-\sigma}^\dagger - c_{i\sigma}^\dagger c_{i-\hat{x}-2\hat{y}, -\sigma}^\dagger -c_{i\sigma} c_{i+\hat{x}+2\hat{y},-\sigma} + c_{i\sigma} c_{i-\hat{x}-2\hat{y}, -\sigma} \nonumber \\
    &- c_{i\sigma}^\dagger c_{i+\hat{x}-2\hat{y},-\sigma}^\dagger + c_{i\sigma}^\dagger c_{i-\hat{x}+2\hat{y}, -\sigma}^\dagger +  c_{i\sigma} c_{i+\hat{x}-2\hat{y},-\sigma} - c_{i\sigma} c_{i-\hat{x}+2\hat{y}, -\sigma}] \nonumber\\
    &+ \frac{\Delta_f}{4} \sum_{i\sigma}[- c_{i\sigma}^\dagger c_{i+2\hat{x}+\hat{y},-\sigma}^\dagger + c_{i\sigma}^\dagger c_{i-2\hat{x}-\hat{y}, -\sigma}^\dagger + c_{i\sigma} c_{i+2\hat{x}+\hat{y},-\sigma} - c_{i\sigma} c_{i-2\hat{x}-\hat{y}, -\sigma} \nonumber \\
    &+c_{i\sigma}^\dagger c_{i+2\hat{x}-\hat{y},-\sigma}^\dagger - c_{i\sigma}^\dagger c_{i-2\hat{x}+\hat{y}, -\sigma}^\dagger-c_{i\sigma} c_{i+2\hat{x}-\hat{y},-\sigma} - c_{i\sigma} c_{i-2\hat{x}+\hat{y}, -\sigma}].
\end{align}
\endgroup
\end{widetext}
Using periodic boundary conditions (PBC) in 2D, an FT gives Eq.~\eqref{eq:BdG} with
\begin{equation}
    \Delta_{\boldsymbol{k}} = \Delta_p \sin k_x +i\Delta_f [\sin (2k_y)\cos(k_x)-2\sin (k_y)\cos(2 k_x)], \label{eq:Deltakrealspace}
\end{equation}
which gives a $p_x + i f_y$-wave gap on the FS for a wide range of chemical potentials. The more idealized functions $\cos(\phi_{\boldsymbol{k}})$ and $\sin(3\phi_{\boldsymbol{k}})$ would require much more complicated real space models. We now consider a ribbon geometry, with PBC in the $x$ direction and open boundary conditions (OBC) in the $y$ direction. Let $c_{i\sigma} \to c_{x_i, y_i, \sigma}$ and introduce a partial FT in the $x$ direction \cite{Bernevig2013},
\begin{equation}
    c_{x_i, y_i,\sigma} = \frac{1}{\sqrt{N_x}} \sum_{k_x} c_{k_x, y_i, \sigma}e^{ik_x x_i}.
\end{equation}
There are $N_x$ sites in the $x$ direction, separated by $a = 1$ such that the 1BZ is $k_x \in [-\pi, \pi)$. Due to PBC, $x_{N_x} = x_{0}$. There are $N_y$ sites in the $y$ direction. With OBC, only $y_0, y_1, \dots y_{N_y-1}$ exist and we let $y_i = i$. Inserting the partial FT, the Hamiltonian can be written as
\begin{equation}
    H = \frac{1}{2} \sum_{k_x} \boldsymbol{c}_{k_x}^\dagger H_{k_x} \boldsymbol{c}_{k_x},
\end{equation}
where
\begin{align}
    \boldsymbol{c}_{k_x}^\dagger &= (c_{k_x, 0,\uparrow}^\dagger, c_{k_x, 0,\downarrow}^\dagger, c_{k_x, 1,\uparrow}^\dagger, \dots, c_{k_x, N_y-1,\downarrow}^\dagger, \nonumber\\
    &c_{-k_x, 0,\uparrow}, c_{-k_x, 0,\downarrow}, c_{-k_x, 1,\uparrow}, \dots, c_{-k_x, N_y-1,\downarrow}),
\end{align}
and $H_{k_x}$ is a $4N_y \times 4N_y$ matrix.

We can now find the bands along $k_x$ by a unitary transformation $U_{k_x}^\dagger H_{k_x} U_{k_x} = D_{k_x}$,
\begin{equation}
    H  = \frac{1}{2} \sum_{k_x} \boldsymbol{d}_{k_x}^\dagger D_{k_x}  \boldsymbol{d}_{k_x},
\end{equation}
where $D_{k_x}$ is diagonal and contains the energy bands.
The diagonalized operators are linear combinations of electron operators at all sites, $\boldsymbol{d}_{k_x} = U_{k_x}^\dagger  \boldsymbol{c}_{k_x}$. The columns of $U_{k_x}$, i.e., the eigenvectors of $H_{k_x}$, give the coefficients in the linear combination. Row $y_i$ enters four places in $\boldsymbol{c}_{k_x}$; in $c_{k_x, y_i,\uparrow}, c_{k_x, y_i,\downarrow}, c_{-k_x, y_i,\uparrow}^\dagger,$ and $c_{-k_x, y_i,\downarrow}^\dagger$. We define $|\psi_i|^2$ as the sum of the absolute squares of the four coefficients in front of these operators, quantifying the weight of each eigenstate on row $y_i$.

Figure \ref{fig:edge} shows the energy bands along $k_x$ revealing zero energy states that are localized at separate edges of the ribbon geometry. The more realistic case of $\Delta_p, \Delta_f \ll t$ gives the same result qualitatively. Larger gap amplitudes are simply convenient for plotting. Additionally, the same qualitative results are found for a wide range of chemical potentials. Figures \ref{fig:edge}(c) and \ref{fig:edge}(d) illustrate the momentum dependence of the gap $\Delta_{\boldsymbol{k}}$ in Eq.~\eqref{eq:Deltakrealspace} showing that the gap is $p_x+if_y$-wave on the FS. 
The appearance of topological edge states establishes that $p_x+if_y$-pairing gives rise to topological superconductivity.

\bibliography{main.bbl}

\begin{thebibliography}{93}%
\makeatletter
\providecommand \@ifxundefined [1]{%
 \@ifx{#1\undefined}
}%
\providecommand \@ifnum [1]{%
 \ifnum #1\expandafter \@firstoftwo
 \else \expandafter \@secondoftwo
 \fi
}%
\providecommand \@ifx [1]{%
 \ifx #1\expandafter \@firstoftwo
 \else \expandafter \@secondoftwo
 \fi
}%
\providecommand \natexlab [1]{#1}%
\providecommand \enquote  [1]{``#1''}%
\providecommand \bibnamefont  [1]{#1}%
\providecommand \bibfnamefont [1]{#1}%
\providecommand \citenamefont [1]{#1}%
\providecommand \href@noop [0]{\@secondoftwo}%
\providecommand \href [0]{\begingroup \@sanitize@url \@href}%
\providecommand \@href[1]{\@@startlink{#1}\@@href}%
\providecommand \@@href[1]{\endgroup#1\@@endlink}%
\providecommand \@sanitize@url [0]{\catcode `\\12\catcode `\$12\catcode `\&12\catcode `\#12\catcode `\^12\catcode `\_12\catcode `\%12\relax}%
\providecommand \@@startlink[1]{}%
\providecommand \@@endlink[0]{}%
\providecommand \url  [0]{\begingroup\@sanitize@url \@url }%
\providecommand \@url [1]{\endgroup\@href {#1}{\urlprefix }}%
\providecommand \urlprefix  [0]{URL }%
\providecommand \Eprint [0]{\href }%
\providecommand \doibase [0]{https://doi.org/}%
\providecommand \selectlanguage [0]{\@gobble}%
\providecommand \bibinfo  [0]{\@secondoftwo}%
\providecommand \bibfield  [0]{\@secondoftwo}%
\providecommand \translation [1]{[#1]}%
\providecommand \BibitemOpen [0]{}%
\providecommand \bibitemStop [0]{}%
\providecommand \bibitemNoStop [0]{.\EOS\space}%
\providecommand \EOS [0]{\spacefactor3000\relax}%
\providecommand \BibitemShut  [1]{\csname bibitem#1\endcsname}%
\let\auto@bib@innerbib\@empty
\bibitem [{\citenamefont {Fossheim}\ and\ \citenamefont {Sudb{\o}}(2004)}]{SC_book}%
  \BibitemOpen
  \bibfield  {author} {\bibinfo {author} {\bibfnamefont {K.}~\bibnamefont {Fossheim}}\ and\ \bibinfo {author} {\bibfnamefont {A.}~\bibnamefont {Sudb{\o}}},\ }\href@noop {} {\emph {\bibinfo {title} {Superconductivity: Physics and Applications}}}\ (\bibinfo  {publisher} {Wiley, Chichester, UK},\ \bibinfo {year} {2004})\BibitemShut {NoStop}%
\bibitem [{\citenamefont {Sigrist}\ and\ \citenamefont {Ueda}(1991)}]{Sigrist1991Apr}%
  \BibitemOpen
  \bibfield  {author} {\bibinfo {author} {\bibfnamefont {M.}~\bibnamefont {Sigrist}}\ and\ \bibinfo {author} {\bibfnamefont {K.}~\bibnamefont {Ueda}},\ }\href {https://doi.org/10.1103/RevModPhys.63.239} {\bibfield  {journal} {\bibinfo  {journal} {Rev. Mod. Phys.}\ }\textbf {\bibinfo {volume} {63}},\ \bibinfo {pages} {239} (\bibinfo {year} {1991})}\BibitemShut {NoStop}%
\bibitem [{\citenamefont {Moore}\ and\ \citenamefont {Smart}(2020)}]{Moore2020Aug}%
  \BibitemOpen
  \bibfield  {author} {\bibinfo {author} {\bibfnamefont {E.~A.}\ \bibnamefont {Moore}}\ and\ \bibinfo {author} {\bibfnamefont {L.~E.}\ \bibnamefont {Smart}},\ }in\ \href {https://doi.org/10.1201/9780429027284-10} {\emph {\bibinfo {booktitle} {{Solid State Chemistry}}}}\ (\bibinfo  {publisher} {CRC Press},\ \bibinfo {address} {Boca Raton, FL, USA},\ \bibinfo {year} {2020})\ pp.\ \bibinfo {pages} {347--362}\BibitemShut {NoStop}%
\bibitem [{\citenamefont {Stewart}(2011)}]{Stewart2011Dec}%
  \BibitemOpen
  \bibfield  {author} {\bibinfo {author} {\bibfnamefont {G.~R.}\ \bibnamefont {Stewart}},\ }\href {https://doi.org/10.1103/RevModPhys.83.1589} {\bibfield  {journal} {\bibinfo  {journal} {Rev. Mod. Phys.}\ }\textbf {\bibinfo {volume} {83}},\ \bibinfo {pages} {1589} (\bibinfo {year} {2011})}\BibitemShut {NoStop}%
\bibitem [{\citenamefont {Onnes}(1911)}]{onnes1911leiden}%
  \BibitemOpen
  \bibfield  {author} {\bibinfo {author} {\bibfnamefont {H.~K.}\ \bibnamefont {Onnes}},\ }\href@noop {} {\bibfield  {journal} {\bibinfo  {journal} {Commun. Phys. Lab. Univ. Leiden 120b, 122b, 124c}\ } (\bibinfo {year} {1911})},\ \bibinfo {note} {reprinted in Proc. K. Ned. Akad. Wet. \textbf{13}, 1274 (1911), \textit{ibid.} \textbf{14}, 113 (1911), \textit{ibid.} \textbf{14}, 818 (1911)}\BibitemShut {NoStop}%
\bibitem [{\citenamefont {van Delft}\ and\ \citenamefont {Kes}(2010)}]{vanDelft2010Sep}%
  \BibitemOpen
  \bibfield  {author} {\bibinfo {author} {\bibfnamefont {D.}~\bibnamefont {van Delft}}\ and\ \bibinfo {author} {\bibfnamefont {P.}~\bibnamefont {Kes}},\ }\href {https://doi.org/10.1063/1.3490499} {\bibfield  {journal} {\bibinfo  {journal} {Phys. Today}\ }\textbf {\bibinfo {volume} {63}},\ \bibinfo {pages} {38} (\bibinfo {year} {2010})}\BibitemShut {NoStop}%
\bibitem [{\citenamefont {Anderson}(1963)}]{Anderson1962}%
  \BibitemOpen
  \bibfield  {author} {\bibinfo {author} {\bibfnamefont {P.~W.}\ \bibnamefont {Anderson}},\ }\href {https://doi.org/10.1103/PhysRev.130.439} {\bibfield  {journal} {\bibinfo  {journal} {Phys. Rev.}\ }\textbf {\bibinfo {volume} {130}},\ \bibinfo {pages} {439} (\bibinfo {year} {1963})}\BibitemShut {NoStop}%
\bibitem [{\citenamefont {Englert}\ and\ \citenamefont {Brout}(1964)}]{Englert1964}%
  \BibitemOpen
  \bibfield  {author} {\bibinfo {author} {\bibfnamefont {F.}~\bibnamefont {Englert}}\ and\ \bibinfo {author} {\bibfnamefont {R.}~\bibnamefont {Brout}},\ }\href {https://doi.org/10.1103/PhysRevLett.13.321} {\bibfield  {journal} {\bibinfo  {journal} {Phys. Rev. Lett.}\ }\textbf {\bibinfo {volume} {13}},\ \bibinfo {pages} {321} (\bibinfo {year} {1964})}\BibitemShut {NoStop}%
\bibitem [{\citenamefont {Higgs}(1964)}]{Higgs1964}%
  \BibitemOpen
  \bibfield  {author} {\bibinfo {author} {\bibfnamefont {P.~W.}\ \bibnamefont {Higgs}},\ }\href {https://doi.org/10.1103/PhysRevLett.13.508} {\bibfield  {journal} {\bibinfo  {journal} {Phys. Rev. Lett.}\ }\textbf {\bibinfo {volume} {13}},\ \bibinfo {pages} {508} (\bibinfo {year} {1964})}\BibitemShut {NoStop}%
\bibitem [{\citenamefont {Maksimov}(2000)}]{Maksimov2000Oct}%
  \BibitemOpen
  \bibfield  {author} {\bibinfo {author} {\bibfnamefont {E.~G.}\ \bibnamefont {Maksimov}},\ }\href {https://doi.org/10.1070/PU2000v043n10ABEH000770} {\bibfield  {journal} {\bibinfo  {journal} {Phys.-Usp.}\ }\textbf {\bibinfo {volume} {43}},\ \bibinfo {pages} {965} (\bibinfo {year} {2000})}\BibitemShut {NoStop}%
\bibitem [{\citenamefont {Nishijima}\ \emph {et~al.}(2013)\citenamefont {Nishijima}, \citenamefont {Eckroad}, \citenamefont {Marian}, \citenamefont {Choi}, \citenamefont {Kim}, \citenamefont {Terai}, \citenamefont {Deng}, \citenamefont {Zheng}, \citenamefont {Wang}, \citenamefont {Umemoto}, \citenamefont {Du}, \citenamefont {Febvre}, \citenamefont {Keenan}, \citenamefont {Mukhanov}, \citenamefont {Cooley}, \citenamefont {Foley}, \citenamefont {Hassenzahl},\ and\ \citenamefont {Izumi}}]{Nishijima2013Sep}%
  \BibitemOpen
  \bibfield  {author} {\bibinfo {author} {\bibfnamefont {S.}~\bibnamefont {Nishijima}}, \bibinfo {author} {\bibfnamefont {S.}~\bibnamefont {Eckroad}}, \bibinfo {author} {\bibfnamefont {A.}~\bibnamefont {Marian}}, \bibinfo {author} {\bibfnamefont {K.}~\bibnamefont {Choi}}, \bibinfo {author} {\bibfnamefont {W.~S.}\ \bibnamefont {Kim}}, \bibinfo {author} {\bibfnamefont {M.}~\bibnamefont {Terai}}, \bibinfo {author} {\bibfnamefont {Z.}~\bibnamefont {Deng}}, \bibinfo {author} {\bibfnamefont {J.}~\bibnamefont {Zheng}}, \bibinfo {author} {\bibfnamefont {J.}~\bibnamefont {Wang}}, \bibinfo {author} {\bibfnamefont {K.}~\bibnamefont {Umemoto}}, \bibinfo {author} {\bibfnamefont {J.}~\bibnamefont {Du}}, \bibinfo {author} {\bibfnamefont {P.}~\bibnamefont {Febvre}}, \bibinfo {author} {\bibfnamefont {S.}~\bibnamefont {Keenan}}, \bibinfo {author} {\bibfnamefont {O.}~\bibnamefont {Mukhanov}}, \bibinfo {author} {\bibfnamefont {L.~D.}\ \bibnamefont {Cooley}}, \bibinfo {author} {\bibfnamefont {C.~P.}\ \bibnamefont {Foley}},
  \bibinfo {author} {\bibfnamefont {W.~V.}\ \bibnamefont {Hassenzahl}},\ and\ \bibinfo {author} {\bibfnamefont {M.}~\bibnamefont {Izumi}},\ }\href {https://doi.org/10.1088/0953-2048/26/11/113001} {\bibfield  {journal} {\bibinfo  {journal} {Supercond. Sci. Technol.}\ }\textbf {\bibinfo {volume} {26}},\ \bibinfo {pages} {113001} (\bibinfo {year} {2013})}\BibitemShut {NoStop}%
\bibitem [{\citenamefont {Hassenzahl}\ \emph {et~al.}(2004)\citenamefont {Hassenzahl}, \citenamefont {Hazelton}, \citenamefont {Johnson}, \citenamefont {Komarek}, \citenamefont {Noe},\ and\ \citenamefont {Reis}}]{Hassenzahl2004Sep}%
  \BibitemOpen
  \bibfield  {author} {\bibinfo {author} {\bibfnamefont {W.~V.}\ \bibnamefont {Hassenzahl}}, \bibinfo {author} {\bibfnamefont {D.~W.}\ \bibnamefont {Hazelton}}, \bibinfo {author} {\bibfnamefont {B.~K.}\ \bibnamefont {Johnson}}, \bibinfo {author} {\bibfnamefont {P.}~\bibnamefont {Komarek}}, \bibinfo {author} {\bibfnamefont {M.}~\bibnamefont {Noe}},\ and\ \bibinfo {author} {\bibfnamefont {C.~T.}\ \bibnamefont {Reis}},\ }\href {https://doi.org/10.1109/JPROC.2004.833674} {\bibfield  {journal} {\bibinfo  {journal} {Proc. IEEE}\ }\textbf {\bibinfo {volume} {92}},\ \bibinfo {pages} {1655} (\bibinfo {year} {2004})}\BibitemShut {NoStop}%
\bibitem [{\citenamefont {Bardeen}\ \emph {et~al.}(1957)\citenamefont {Bardeen}, \citenamefont {Cooper},\ and\ \citenamefont {Schrieffer}}]{BCS}%
  \BibitemOpen
  \bibfield  {author} {\bibinfo {author} {\bibfnamefont {J.}~\bibnamefont {Bardeen}}, \bibinfo {author} {\bibfnamefont {L.~N.}\ \bibnamefont {Cooper}},\ and\ \bibinfo {author} {\bibfnamefont {J.~R.}\ \bibnamefont {Schrieffer}},\ }\href {https://doi.org/10.1103/PhysRev.108.1175} {\bibfield  {journal} {\bibinfo  {journal} {Phys. Rev.}\ }\textbf {\bibinfo {volume} {108}},\ \bibinfo {pages} {1175} (\bibinfo {year} {1957})}\BibitemShut {NoStop}%
\bibitem [{\citenamefont {Chumak}\ \emph {et~al.}(2015)\citenamefont {Chumak}, \citenamefont {Vasyuchka}, \citenamefont {Serga},\ and\ \citenamefont {Hillebrands}}]{Chumak2015Jun}%
  \BibitemOpen
  \bibfield  {author} {\bibinfo {author} {\bibfnamefont {A.~V.}\ \bibnamefont {Chumak}}, \bibinfo {author} {\bibfnamefont {V.~I.}\ \bibnamefont {Vasyuchka}}, \bibinfo {author} {\bibfnamefont {A.~A.}\ \bibnamefont {Serga}},\ and\ \bibinfo {author} {\bibfnamefont {B.}~\bibnamefont {Hillebrands}},\ }\href {https://doi.org/10.1038/nphys3347} {\bibfield  {journal} {\bibinfo  {journal} {Nat. Phys.}\ }\textbf {\bibinfo {volume} {11}},\ \bibinfo {pages} {453} (\bibinfo {year} {2015})}\BibitemShut {NoStop}%
\bibitem [{\citenamefont {Brataas}\ \emph {et~al.}(2020)\citenamefont {Brataas}, \citenamefont {van Wees}, \citenamefont {Klein}, \citenamefont {de~Loubens},\ and\ \citenamefont {Viret}}]{Brataas2020Nov}%
  \BibitemOpen
  \bibfield  {author} {\bibinfo {author} {\bibfnamefont {A.}~\bibnamefont {Brataas}}, \bibinfo {author} {\bibfnamefont {B.}~\bibnamefont {van Wees}}, \bibinfo {author} {\bibfnamefont {O.}~\bibnamefont {Klein}}, \bibinfo {author} {\bibfnamefont {G.}~\bibnamefont {de~Loubens}},\ and\ \bibinfo {author} {\bibfnamefont {M.}~\bibnamefont {Viret}},\ }\href {https://doi.org/10.1016/j.physrep.2020.08.006} {\bibfield  {journal} {\bibinfo  {journal} {Phys. Rep.}\ }\textbf {\bibinfo {volume} {885}},\ \bibinfo {pages} {1} (\bibinfo {year} {2020})}\BibitemShut {NoStop}%
\bibitem [{\citenamefont {Monthoux}\ and\ \citenamefont {Pines}(1993)}]{Pines1993}%
  \BibitemOpen
  \bibfield  {author} {\bibinfo {author} {\bibfnamefont {P.}~\bibnamefont {Monthoux}}\ and\ \bibinfo {author} {\bibfnamefont {D.}~\bibnamefont {Pines}},\ }\href {https://doi.org/10.1103/PhysRevB.47.6069} {\bibfield  {journal} {\bibinfo  {journal} {Phys. Rev. B}\ }\textbf {\bibinfo {volume} {47}},\ \bibinfo {pages} {6069} (\bibinfo {year} {1993})}\BibitemShut {NoStop}%
\bibitem [{\citenamefont {Scalapino}(1999)}]{SCspinHistory_Scalapino1999}%
  \BibitemOpen
  \bibfield  {author} {\bibinfo {author} {\bibfnamefont {D.~J.}\ \bibnamefont {Scalapino}},\ }\href {https://doi.org/10.1023/A:1022559920049} {\bibfield  {journal} {\bibinfo  {journal} {J. Low Temp. Phys.}\ }\textbf {\bibinfo {volume} {117}},\ \bibinfo {pages} {179} (\bibinfo {year} {1999})}\BibitemShut {NoStop}%
\bibitem [{\citenamefont {Moriya}\ and\ \citenamefont {Ueda}(2003)}]{SCAFMspinMoriya2003Jul}%
  \BibitemOpen
  \bibfield  {author} {\bibinfo {author} {\bibfnamefont {T.}~\bibnamefont {Moriya}}\ and\ \bibinfo {author} {\bibfnamefont {K.}~\bibnamefont {Ueda}},\ }\href {https://doi.org/10.1088/0034-4885/66/8/202} {\bibfield  {journal} {\bibinfo  {journal} {Rep. Prog. Phys.}\ }\textbf {\bibinfo {volume} {66}},\ \bibinfo {pages} {1299} (\bibinfo {year} {2003})}\BibitemShut {NoStop}%
\bibitem [{\citenamefont {Hirschfeld}\ \emph {et~al.}(2011)\citenamefont {Hirschfeld}, \citenamefont {Korshunov},\ and\ \citenamefont {Mazin}}]{SCspinGapSym_Hirschfeld2011}%
  \BibitemOpen
  \bibfield  {author} {\bibinfo {author} {\bibfnamefont {P.~J.}\ \bibnamefont {Hirschfeld}}, \bibinfo {author} {\bibfnamefont {M.~M.}\ \bibnamefont {Korshunov}},\ and\ \bibinfo {author} {\bibfnamefont {I.~I.}\ \bibnamefont {Mazin}},\ }\href {https://doi.org/10.1088/0034-4885/74/12/124508} {\bibfield  {journal} {\bibinfo  {journal} {Rep. Prog. Phys.}\ }\textbf {\bibinfo {volume} {74}},\ \bibinfo {pages} {124508} (\bibinfo {year} {2011})}\BibitemShut {NoStop}%
\bibitem [{\citenamefont {Stewart}(1984)}]{HeavyFermionReview1984}%
  \BibitemOpen
  \bibfield  {author} {\bibinfo {author} {\bibfnamefont {G.~R.}\ \bibnamefont {Stewart}},\ }\href {https://doi.org/10.1103/RevModPhys.56.755} {\bibfield  {journal} {\bibinfo  {journal} {Rev. Mod. Phys.}\ }\textbf {\bibinfo {volume} {56}},\ \bibinfo {pages} {755} (\bibinfo {year} {1984})}\BibitemShut {NoStop}%
\bibitem [{\citenamefont {Wirth}\ and\ \citenamefont {Steglich}(2016)}]{NatRevMater2016}%
  \BibitemOpen
  \bibfield  {author} {\bibinfo {author} {\bibfnamefont {S.}~\bibnamefont {Wirth}}\ and\ \bibinfo {author} {\bibfnamefont {F.}~\bibnamefont {Steglich}},\ }\href {https://doi.org/10.1038/natrevmats.2016.51} {\bibfield  {journal} {\bibinfo  {journal} {Nat. Rev. Mater.}\ }\textbf {\bibinfo {volume} {1}},\ \bibinfo {pages} {16051} (\bibinfo {year} {2016})}\BibitemShut {NoStop}%
\bibitem [{\citenamefont {Jiao}\ \emph {et~al.}(2020)\citenamefont {Jiao}, \citenamefont {Howard}, \citenamefont {Ran}, \citenamefont {Wang}, \citenamefont {Rodriguez}, \citenamefont {Sigrist}, \citenamefont {Wang}, \citenamefont {Butch},\ and\ \citenamefont {Madhavan}}]{NatureChiralSC2020}%
  \BibitemOpen
  \bibfield  {author} {\bibinfo {author} {\bibfnamefont {L.}~\bibnamefont {Jiao}}, \bibinfo {author} {\bibfnamefont {S.}~\bibnamefont {Howard}}, \bibinfo {author} {\bibfnamefont {S.}~\bibnamefont {Ran}}, \bibinfo {author} {\bibfnamefont {Z.}~\bibnamefont {Wang}}, \bibinfo {author} {\bibfnamefont {J.~O.}\ \bibnamefont {Rodriguez}}, \bibinfo {author} {\bibfnamefont {M.}~\bibnamefont {Sigrist}}, \bibinfo {author} {\bibfnamefont {Z.}~\bibnamefont {Wang}}, \bibinfo {author} {\bibfnamefont {N.~P.}\ \bibnamefont {Butch}},\ and\ \bibinfo {author} {\bibfnamefont {V.}~\bibnamefont {Madhavan}},\ }\href {https://doi.org/10.1038/s41586-020-2122-2} {\bibfield  {journal} {\bibinfo  {journal} {Nature}\ }\textbf {\bibinfo {volume} {579}},\ \bibinfo {pages} {523} (\bibinfo {year} {2020})}\BibitemShut {NoStop}%
\bibitem [{\citenamefont {Rohling}\ \emph {et~al.}(2018)\citenamefont {Rohling}, \citenamefont {Fj{\ae}rbu},\ and\ \citenamefont {Brataas}}]{Rohling2018Mar}%
  \BibitemOpen
  \bibfield  {author} {\bibinfo {author} {\bibfnamefont {N.}~\bibnamefont {Rohling}}, \bibinfo {author} {\bibfnamefont {E.~L.}\ \bibnamefont {Fj{\ae}rbu}},\ and\ \bibinfo {author} {\bibfnamefont {A.}~\bibnamefont {Brataas}},\ }\href {https://doi.org/10.1103/PhysRevB.97.115401} {\bibfield  {journal} {\bibinfo  {journal} {Phys. Rev. B}\ }\textbf {\bibinfo {volume} {97}},\ \bibinfo {pages} {115401} (\bibinfo {year} {2018})}\BibitemShut {NoStop}%
\bibitem [{\citenamefont {Fj{\ae}rbu}\ \emph {et~al.}(2019)\citenamefont {Fj{\ae}rbu}, \citenamefont {Rohling},\ and\ \citenamefont {Brataas}}]{Fjaerbu2019Sep}%
  \BibitemOpen
  \bibfield  {author} {\bibinfo {author} {\bibfnamefont {E.~L.}\ \bibnamefont {Fj{\ae}rbu}}, \bibinfo {author} {\bibfnamefont {N.}~\bibnamefont {Rohling}},\ and\ \bibinfo {author} {\bibfnamefont {A.}~\bibnamefont {Brataas}},\ }\href {https://doi.org/10.1103/PhysRevB.100.125432} {\bibfield  {journal} {\bibinfo  {journal} {Phys. Rev. B}\ }\textbf {\bibinfo {volume} {100}},\ \bibinfo {pages} {125432} (\bibinfo {year} {2019})}\BibitemShut {NoStop}%
\bibitem [{\citenamefont {Erlandsen}\ \emph {et~al.}(2019)\citenamefont {Erlandsen}, \citenamefont {Kamra}, \citenamefont {Brataas},\ and\ \citenamefont {Sudb{\o}}}]{Erlandsen2019Sep}%
  \BibitemOpen
  \bibfield  {author} {\bibinfo {author} {\bibfnamefont {E.}~\bibnamefont {Erlandsen}}, \bibinfo {author} {\bibfnamefont {A.}~\bibnamefont {Kamra}}, \bibinfo {author} {\bibfnamefont {A.}~\bibnamefont {Brataas}},\ and\ \bibinfo {author} {\bibfnamefont {A.}~\bibnamefont {Sudb{\o}}},\ }\href {https://doi.org/10.1103/PhysRevB.100.100503} {\bibfield  {journal} {\bibinfo  {journal} {Phys. Rev. B}\ }\textbf {\bibinfo {volume} {100}},\ \bibinfo {pages} {100503} (\bibinfo {year} {2019})}\BibitemShut {NoStop}%
\bibitem [{\citenamefont {Thingstad}\ \emph {et~al.}(2021)\citenamefont {Thingstad}, \citenamefont {Erlandsen},\ and\ \citenamefont {Sudb{\o}}}]{Eliash_Even}%
  \BibitemOpen
  \bibfield  {author} {\bibinfo {author} {\bibfnamefont {E.}~\bibnamefont {Thingstad}}, \bibinfo {author} {\bibfnamefont {E.}~\bibnamefont {Erlandsen}},\ and\ \bibinfo {author} {\bibfnamefont {A.}~\bibnamefont {Sudb{\o}}},\ }\href {https://doi.org/10.1103/PhysRevB.104.014508} {\bibfield  {journal} {\bibinfo  {journal} {Phys. Rev. B}\ }\textbf {\bibinfo {volume} {104}},\ \bibinfo {pages} {014508} (\bibinfo {year} {2021})}\BibitemShut {NoStop}%
\bibitem [{\citenamefont {Sun}\ \emph {et~al.}(2023)\citenamefont {Sun}, \citenamefont {M{\ae}land},\ and\ \citenamefont {Sudb{\o}}}]{Sun2023Aug}%
  \BibitemOpen
  \bibfield  {author} {\bibinfo {author} {\bibfnamefont {C.}~\bibnamefont {Sun}}, \bibinfo {author} {\bibfnamefont {K.}~\bibnamefont {M{\ae}land}},\ and\ \bibinfo {author} {\bibfnamefont {A.}~\bibnamefont {Sudb{\o}}},\ }\href {https://doi.org/10.1103/PhysRevB.108.054520} {\bibfield  {journal} {\bibinfo  {journal} {Phys. Rev. B}\ }\textbf {\bibinfo {volume} {108}},\ \bibinfo {pages} {054520} (\bibinfo {year} {2023})}\BibitemShut {NoStop}%
\bibitem [{\citenamefont {M{\ae}land}\ and\ \citenamefont {Sudb{\o}}(2023{\natexlab{a}})}]{Maeland2023Apr}%
  \BibitemOpen
  \bibfield  {author} {\bibinfo {author} {\bibfnamefont {K.}~\bibnamefont {M{\ae}land}}\ and\ \bibinfo {author} {\bibfnamefont {A.}~\bibnamefont {Sudb{\o}}},\ }\href {https://doi.org/10.1103/PhysRevLett.130.156002} {\bibfield  {journal} {\bibinfo  {journal} {Phys. Rev. Lett.}\ }\textbf {\bibinfo {volume} {130}},\ \bibinfo {pages} {156002} (\bibinfo {year} {2023}{\natexlab{a}})}\BibitemShut {NoStop}%
\bibitem [{\citenamefont {M{\ae}land}\ \emph {et~al.}(2023)\citenamefont {M{\ae}land}, \citenamefont {Abnar}, \citenamefont {Benestad},\ and\ \citenamefont {Sudb{\o}}}]{Maeland2023Dec}%
  \BibitemOpen
  \bibfield  {author} {\bibinfo {author} {\bibfnamefont {K.}~\bibnamefont {M{\ae}land}}, \bibinfo {author} {\bibfnamefont {S.}~\bibnamefont {Abnar}}, \bibinfo {author} {\bibfnamefont {J.}~\bibnamefont {Benestad}},\ and\ \bibinfo {author} {\bibfnamefont {A.}~\bibnamefont {Sudb{\o}}},\ }\href {https://doi.org/10.1103/PhysRevB.108.224515} {\bibfield  {journal} {\bibinfo  {journal} {Phys. Rev. B}\ }\textbf {\bibinfo {volume} {108}},\ \bibinfo {pages} {224515} (\bibinfo {year} {2023})}\BibitemShut {NoStop}%
\bibitem [{\citenamefont {Bostr{\ifmmode\ddot{o}\else\"{o}\fi}m}\ and\ \citenamefont {Bostr{\ifmmode\ddot{o}\else\"{o}\fi}m}(2023)}]{Bostrom2023Dec}%
  \BibitemOpen
  \bibfield  {author} {\bibinfo {author} {\bibfnamefont {F.~V.}\ \bibnamefont {Bostr{\ifmmode\ddot{o}\else\"{o}\fi}m}}\ and\ \bibinfo {author} {\bibfnamefont {E.~V.}\ \bibnamefont {Bostr{\ifmmode\ddot{o}\else\"{o}\fi}m}},\ }\href {https://arxiv.org/abs/2312.02655v1} {\bibfield  {journal} {\bibinfo  {journal} {arXiv:2312.02655}\ } (\bibinfo {year} {2023})}\BibitemShut {NoStop}%
\bibitem [{\citenamefont {Kajiwara}\ \emph {et~al.}(2010)\citenamefont {Kajiwara}, \citenamefont {Harii}, \citenamefont {Takahashi}, \citenamefont {Ohe}, \citenamefont {Uchida}, \citenamefont {Mizuguchi}, \citenamefont {Umezawa}, \citenamefont {Kawai}, \citenamefont {Ando}, \citenamefont {Takanashi}, \citenamefont {Maekawa},\ and\ \citenamefont {Saitoh}}]{Kajiwara2010NMMIexp}%
  \BibitemOpen
  \bibfield  {author} {\bibinfo {author} {\bibfnamefont {Y.}~\bibnamefont {Kajiwara}}, \bibinfo {author} {\bibfnamefont {K.}~\bibnamefont {Harii}}, \bibinfo {author} {\bibfnamefont {S.}~\bibnamefont {Takahashi}}, \bibinfo {author} {\bibfnamefont {J.}~\bibnamefont {Ohe}}, \bibinfo {author} {\bibfnamefont {K.}~\bibnamefont {Uchida}}, \bibinfo {author} {\bibfnamefont {M.}~\bibnamefont {Mizuguchi}}, \bibinfo {author} {\bibfnamefont {H.}~\bibnamefont {Umezawa}}, \bibinfo {author} {\bibfnamefont {H.}~\bibnamefont {Kawai}}, \bibinfo {author} {\bibfnamefont {K.}~\bibnamefont {Ando}}, \bibinfo {author} {\bibfnamefont {K.}~\bibnamefont {Takanashi}}, \bibinfo {author} {\bibfnamefont {S.}~\bibnamefont {Maekawa}},\ and\ \bibinfo {author} {\bibfnamefont {E.}~\bibnamefont {Saitoh}},\ }\href {https://doi.org/10.1038/nature08876} {\bibfield  {journal} {\bibinfo  {journal} {Nature}\ }\textbf {\bibinfo {volume} {464}},\ \bibinfo {pages} {262} (\bibinfo {year} {2010})}\BibitemShut {NoStop}%
\bibitem [{\citenamefont {Li}\ \emph {et~al.}(2016)\citenamefont {Li}, \citenamefont {Xu}, \citenamefont {Aldosary}, \citenamefont {Tang}, \citenamefont {Lin}, \citenamefont {Zhang}, \citenamefont {Lake},\ and\ \citenamefont {Shi}}]{Li2016NMMIexp}%
  \BibitemOpen
  \bibfield  {author} {\bibinfo {author} {\bibfnamefont {J.}~\bibnamefont {Li}}, \bibinfo {author} {\bibfnamefont {Y.}~\bibnamefont {Xu}}, \bibinfo {author} {\bibfnamefont {M.}~\bibnamefont {Aldosary}}, \bibinfo {author} {\bibfnamefont {C.}~\bibnamefont {Tang}}, \bibinfo {author} {\bibfnamefont {Z.}~\bibnamefont {Lin}}, \bibinfo {author} {\bibfnamefont {S.}~\bibnamefont {Zhang}}, \bibinfo {author} {\bibfnamefont {R.}~\bibnamefont {Lake}},\ and\ \bibinfo {author} {\bibfnamefont {J.}~\bibnamefont {Shi}},\ }\href {https://doi.org/10.1038/ncomms10858} {\bibfield  {journal} {\bibinfo  {journal} {Nat. Commun.}\ }\textbf {\bibinfo {volume} {7}},\ \bibinfo {pages} {10858} (\bibinfo {year} {2016})}\BibitemShut {NoStop}%
\bibitem [{\citenamefont {Wu}\ \emph {et~al.}(2016)\citenamefont {Wu}, \citenamefont {Wan}, \citenamefont {Zhang}, \citenamefont {Yuan}, \citenamefont {Zhang}, \citenamefont {Qin}, \citenamefont {Wei}, \citenamefont {Han},\ and\ \citenamefont {Zhang}}]{Wu2016NMMIexp}%
  \BibitemOpen
  \bibfield  {author} {\bibinfo {author} {\bibfnamefont {H.}~\bibnamefont {Wu}}, \bibinfo {author} {\bibfnamefont {C.~H.}\ \bibnamefont {Wan}}, \bibinfo {author} {\bibfnamefont {X.}~\bibnamefont {Zhang}}, \bibinfo {author} {\bibfnamefont {Z.~H.}\ \bibnamefont {Yuan}}, \bibinfo {author} {\bibfnamefont {Q.~T.}\ \bibnamefont {Zhang}}, \bibinfo {author} {\bibfnamefont {J.~Y.}\ \bibnamefont {Qin}}, \bibinfo {author} {\bibfnamefont {H.~X.}\ \bibnamefont {Wei}}, \bibinfo {author} {\bibfnamefont {X.~F.}\ \bibnamefont {Han}},\ and\ \bibinfo {author} {\bibfnamefont {S.}~\bibnamefont {Zhang}},\ }\href {https://doi.org/10.1103/PhysRevB.93.060403} {\bibfield  {journal} {\bibinfo  {journal} {Phys. Rev. B}\ }\textbf {\bibinfo {volume} {93}},\ \bibinfo {pages} {060403} (\bibinfo {year} {2016})}\BibitemShut {NoStop}%
\bibitem [{\citenamefont {Kargarian}\ \emph {et~al.}(2016)\citenamefont {Kargarian}, \citenamefont {Efimkin},\ and\ \citenamefont {Galitski}}]{KargarianFMTI}%
  \BibitemOpen
  \bibfield  {author} {\bibinfo {author} {\bibfnamefont {M.}~\bibnamefont {Kargarian}}, \bibinfo {author} {\bibfnamefont {D.~K.}\ \bibnamefont {Efimkin}},\ and\ \bibinfo {author} {\bibfnamefont {V.}~\bibnamefont {Galitski}},\ }\href {https://doi.org/10.1103/PhysRevLett.117.076806} {\bibfield  {journal} {\bibinfo  {journal} {Phys. Rev. Lett.}\ }\textbf {\bibinfo {volume} {117}},\ \bibinfo {pages} {076806} (\bibinfo {year} {2016})}\BibitemShut {NoStop}%
\bibitem [{\citenamefont {Hugdal}\ \emph {et~al.}(2018)\citenamefont {Hugdal}, \citenamefont {Rex}, \citenamefont {Nogueira},\ and\ \citenamefont {Sudb{\o}}}]{Hugdal2018May}%
  \BibitemOpen
  \bibfield  {author} {\bibinfo {author} {\bibfnamefont {H.~G.}\ \bibnamefont {Hugdal}}, \bibinfo {author} {\bibfnamefont {S.}~\bibnamefont {Rex}}, \bibinfo {author} {\bibfnamefont {F.~S.}\ \bibnamefont {Nogueira}},\ and\ \bibinfo {author} {\bibfnamefont {A.}~\bibnamefont {Sudb{\o}}},\ }\href {https://doi.org/10.1103/PhysRevB.97.195438} {\bibfield  {journal} {\bibinfo  {journal} {Phys. Rev. B}\ }\textbf {\bibinfo {volume} {97}},\ \bibinfo {pages} {195438} (\bibinfo {year} {2018})}\BibitemShut {NoStop}%
\bibitem [{\citenamefont {Hugdal}\ and\ \citenamefont {Sudb{\o}}(2020)}]{Hugdal2020FMTI}%
  \BibitemOpen
  \bibfield  {author} {\bibinfo {author} {\bibfnamefont {H.~G.}\ \bibnamefont {Hugdal}}\ and\ \bibinfo {author} {\bibfnamefont {A.}~\bibnamefont {Sudb{\o}}},\ }\href {https://doi.org/10.1103/PhysRevB.102.125429} {\bibfield  {journal} {\bibinfo  {journal} {Phys. Rev. B}\ }\textbf {\bibinfo {volume} {102}},\ \bibinfo {pages} {125429} (\bibinfo {year} {2020})}\BibitemShut {NoStop}%
\bibitem [{\citenamefont {Erlandsen}\ \emph {et~al.}(2020)\citenamefont {Erlandsen}, \citenamefont {Brataas},\ and\ \citenamefont {Sudb{\o}}}]{Erlandsen2020Mar}%
  \BibitemOpen
  \bibfield  {author} {\bibinfo {author} {\bibfnamefont {E.}~\bibnamefont {Erlandsen}}, \bibinfo {author} {\bibfnamefont {A.}~\bibnamefont {Brataas}},\ and\ \bibinfo {author} {\bibfnamefont {A.}~\bibnamefont {Sudb{\o}}},\ }\href {https://doi.org/10.1103/PhysRevB.101.094503} {\bibfield  {journal} {\bibinfo  {journal} {Phys. Rev. B}\ }\textbf {\bibinfo {volume} {101}},\ \bibinfo {pages} {094503} (\bibinfo {year} {2020})}\BibitemShut {NoStop}%
\bibitem [{\citenamefont {Jungwirth}\ \emph {et~al.}(2016)\citenamefont {Jungwirth}, \citenamefont {Marti}, \citenamefont {Wadley},\ and\ \citenamefont {Wunderlich}}]{Jungwirth2016Mar}%
  \BibitemOpen
  \bibfield  {author} {\bibinfo {author} {\bibfnamefont {T.}~\bibnamefont {Jungwirth}}, \bibinfo {author} {\bibfnamefont {X.}~\bibnamefont {Marti}}, \bibinfo {author} {\bibfnamefont {P.}~\bibnamefont {Wadley}},\ and\ \bibinfo {author} {\bibfnamefont {J.}~\bibnamefont {Wunderlich}},\ }\href {https://doi.org/10.1038/nnano.2016.18} {\bibfield  {journal} {\bibinfo  {journal} {Nat. Nanotechnol.}\ }\textbf {\bibinfo {volume} {11}},\ \bibinfo {pages} {231} (\bibinfo {year} {2016})}\BibitemShut {NoStop}%
\bibitem [{\citenamefont {Sun}\ \emph {et~al.}(2022)\citenamefont {Sun}, \citenamefont {Yang},\ and\ \citenamefont {Jalil}}]{Sun2022Mar}%
  \BibitemOpen
  \bibfield  {author} {\bibinfo {author} {\bibfnamefont {C.}~\bibnamefont {Sun}}, \bibinfo {author} {\bibfnamefont {H.}~\bibnamefont {Yang}},\ and\ \bibinfo {author} {\bibfnamefont {M.~B.~A.}\ \bibnamefont {Jalil}},\ }\href {https://doi.org/10.1103/PhysRevB.105.104407} {\bibfield  {journal} {\bibinfo  {journal} {Phys. Rev. B}\ }\textbf {\bibinfo {volume} {105}},\ \bibinfo {pages} {104407} (\bibinfo {year} {2022})}\BibitemShut {NoStop}%
\bibitem [{\citenamefont {Nogués}\ \emph {et~al.}(2005)\citenamefont {Nogués}, \citenamefont {Sort}, \citenamefont {Langlais}, \citenamefont {Skumryev}, \citenamefont {Suriñach}, \citenamefont {Muñoz},\ and\ \citenamefont {Baró}}]{Nogues2005}%
  \BibitemOpen
  \bibfield  {author} {\bibinfo {author} {\bibfnamefont {J.}~\bibnamefont {Nogués}}, \bibinfo {author} {\bibfnamefont {J.}~\bibnamefont {Sort}}, \bibinfo {author} {\bibfnamefont {V.}~\bibnamefont {Langlais}}, \bibinfo {author} {\bibfnamefont {V.}~\bibnamefont {Skumryev}}, \bibinfo {author} {\bibfnamefont {S.}~\bibnamefont {Suriñach}}, \bibinfo {author} {\bibfnamefont {J.}~\bibnamefont {Muñoz}},\ and\ \bibinfo {author} {\bibfnamefont {M.}~\bibnamefont {Baró}},\ }\href {https://doi.org/10.1016/j.physrep.2005.08.004} {\bibfield  {journal} {\bibinfo  {journal} {Phys. Rep.}\ }\textbf {\bibinfo {volume} {422}},\ \bibinfo {pages} {65–117} (\bibinfo {year} {2005})}\BibitemShut {NoStop}%
\bibitem [{\citenamefont {Wu}\ \emph {et~al.}(2020)\citenamefont {Wu}, \citenamefont {Yin}, \citenamefont {Pan}, \citenamefont {Grutter}, \citenamefont {Pan}, \citenamefont {Lee}, \citenamefont {Gilbert}, \citenamefont {Borchers}, \citenamefont {Ratcliff}, \citenamefont {Li}, \citenamefont {Han},\ and\ \citenamefont {Wang}}]{Wu2020}%
  \BibitemOpen
  \bibfield  {author} {\bibinfo {author} {\bibfnamefont {Y.}~\bibnamefont {Wu}}, \bibinfo {author} {\bibfnamefont {G.}~\bibnamefont {Yin}}, \bibinfo {author} {\bibfnamefont {L.}~\bibnamefont {Pan}}, \bibinfo {author} {\bibfnamefont {A.~J.}\ \bibnamefont {Grutter}}, \bibinfo {author} {\bibfnamefont {Q.}~\bibnamefont {Pan}}, \bibinfo {author} {\bibfnamefont {A.}~\bibnamefont {Lee}}, \bibinfo {author} {\bibfnamefont {D.~A.}\ \bibnamefont {Gilbert}}, \bibinfo {author} {\bibfnamefont {J.~A.}\ \bibnamefont {Borchers}}, \bibinfo {author} {\bibfnamefont {W.}~\bibnamefont {Ratcliff}}, \bibinfo {author} {\bibfnamefont {A.}~\bibnamefont {Li}}, \bibinfo {author} {\bibfnamefont {X.-d.}\ \bibnamefont {Han}},\ and\ \bibinfo {author} {\bibfnamefont {K.~L.}\ \bibnamefont {Wang}},\ }\href {https://doi.org/10.1038/s41928-020-0458-0} {\bibfield  {journal} {\bibinfo  {journal} {Nat. Electron.}\ }\textbf {\bibinfo {volume} {3}},\ \bibinfo {pages} {604–611} (\bibinfo {year} {2020})}\BibitemShut {NoStop}%
\bibitem [{\citenamefont {Kamra}\ \emph {et~al.}(2019)\citenamefont {Kamra}, \citenamefont {Thingstad}, \citenamefont {Rastelli}, \citenamefont {Duine}, \citenamefont {Brataas}, \citenamefont {Belzig},\ and\ \citenamefont {Sudb{\o}}}]{Kamra2019afmsqueezed}%
  \BibitemOpen
  \bibfield  {author} {\bibinfo {author} {\bibfnamefont {A.}~\bibnamefont {Kamra}}, \bibinfo {author} {\bibfnamefont {E.}~\bibnamefont {Thingstad}}, \bibinfo {author} {\bibfnamefont {G.}~\bibnamefont {Rastelli}}, \bibinfo {author} {\bibfnamefont {R.~A.}\ \bibnamefont {Duine}}, \bibinfo {author} {\bibfnamefont {A.}~\bibnamefont {Brataas}}, \bibinfo {author} {\bibfnamefont {W.}~\bibnamefont {Belzig}},\ and\ \bibinfo {author} {\bibfnamefont {A.}~\bibnamefont {Sudb{\o}}},\ }\href {https://doi.org/10.1103/PhysRevB.100.174407} {\bibfield  {journal} {\bibinfo  {journal} {Phys. Rev. B}\ }\textbf {\bibinfo {volume} {100}},\ \bibinfo {pages} {174407} (\bibinfo {year} {2019})}\BibitemShut {NoStop}%
\bibitem [{\citenamefont {Hutchinson}\ and\ \citenamefont {Marsiglio}(2020)}]{Hutchinson2020Nov}%
  \BibitemOpen
  \bibfield  {author} {\bibinfo {author} {\bibfnamefont {J.}~\bibnamefont {Hutchinson}}\ and\ \bibinfo {author} {\bibfnamefont {F.}~\bibnamefont {Marsiglio}},\ }\href {https://doi.org/10.1088/1361-648X/abc801} {\bibfield  {journal} {\bibinfo  {journal} {J. Phys.: Condens. Matter}\ }\textbf {\bibinfo {volume} {33}},\ \bibinfo {pages} {065603} (\bibinfo {year} {2020})}\BibitemShut {NoStop}%
\bibitem [{\citenamefont {Eliashberg}(1960{\natexlab{a}})}]{Eliashberg1960Sep}%
  \BibitemOpen
  \bibfield  {author} {\bibinfo {author} {\bibfnamefont {G.~M.}\ \bibnamefont {Eliashberg}},\ }\href@noop {} {\bibfield  {journal} {\bibinfo  {journal} {Zh. Eksp. Teor. Fiz.}\ }\textbf {\bibinfo {volume} {38}},\ \bibinfo {pages} {966} (\bibinfo {year} {1960}{\natexlab{a}})},\ \bibinfo {note} {[Sov. Phys. JETP \textbf{11}, 696 (1960)]}\BibitemShut {NoStop}%
\bibitem [{\citenamefont {Eliashberg}(1960{\natexlab{b}})}]{Eliashberg1961}%
  \BibitemOpen
  \bibfield  {author} {\bibinfo {author} {\bibfnamefont {G.~M.}\ \bibnamefont {Eliashberg}},\ }\href@noop {} {\bibfield  {journal} {\bibinfo  {journal} {Zh. Eksp. Teor. Fiz.}\ }\textbf {\bibinfo {volume} {39}},\ \bibinfo {pages} {1437} (\bibinfo {year} {1960}{\natexlab{b}})},\ \bibinfo {note} {[Sov. Phys. JETP \textbf{12}, 1000 (1961)]}\BibitemShut {NoStop}%
\bibitem [{\citenamefont {Carbotte}(1990)}]{Carbotte1990FreeEnergy}%
  \BibitemOpen
  \bibfield  {author} {\bibinfo {author} {\bibfnamefont {J.~P.}\ \bibnamefont {Carbotte}},\ }\href {https://doi.org/10.1103/RevModPhys.62.1027} {\bibfield  {journal} {\bibinfo  {journal} {Rev. Mod. Phys.}\ }\textbf {\bibinfo {volume} {62}},\ \bibinfo {pages} {1027} (\bibinfo {year} {1990})}\BibitemShut {NoStop}%
\bibitem [{\citenamefont {Marsiglio}(2020)}]{EliashbergRevMarsiglio2020}%
  \BibitemOpen
  \bibfield  {author} {\bibinfo {author} {\bibfnamefont {F.}~\bibnamefont {Marsiglio}},\ }\href {https://doi.org/10.1016/j.aop.2020.168102} {\bibfield  {journal} {\bibinfo  {journal} {Ann. Phys.}\ }\textbf {\bibinfo {volume} {417}},\ \bibinfo {pages} {168102} (\bibinfo {year} {2020})}\BibitemShut {NoStop}%
\bibitem [{\citenamefont {Chubukov}\ \emph {et~al.}(2020)\citenamefont {Chubukov}, \citenamefont {Abanov}, \citenamefont {Esterlis},\ and\ \citenamefont {Kivelson}}]{Chubukov2020FEspecialized}%
  \BibitemOpen
  \bibfield  {author} {\bibinfo {author} {\bibfnamefont {A.~V.}\ \bibnamefont {Chubukov}}, \bibinfo {author} {\bibfnamefont {A.}~\bibnamefont {Abanov}}, \bibinfo {author} {\bibfnamefont {I.}~\bibnamefont {Esterlis}},\ and\ \bibinfo {author} {\bibfnamefont {S.~A.}\ \bibnamefont {Kivelson}},\ }\href {https://doi.org/10.1016/j.aop.2020.168190} {\bibfield  {journal} {\bibinfo  {journal} {Ann. Phys.}\ }\textbf {\bibinfo {volume} {417}},\ \bibinfo {pages} {168190} (\bibinfo {year} {2020})}\BibitemShut {NoStop}%
\bibitem [{\citenamefont {Sigrist}(2000)}]{Sigrist2000TRSB}%
  \BibitemOpen
  \bibfield  {author} {\bibinfo {author} {\bibfnamefont {M.}~\bibnamefont {Sigrist}},\ }\href {https://doi.org/10.1016/S0921-4526(99)01547-1} {\bibfield  {journal} {\bibinfo  {journal} {Physica B}\ }\textbf {\bibinfo {volume} {280}},\ \bibinfo {pages} {154} (\bibinfo {year} {2000})}\BibitemShut {NoStop}%
\bibitem [{\citenamefont {Fernandes}\ \emph {et~al.}(2022)\citenamefont {Fernandes}, \citenamefont {Coldea}, \citenamefont {Ding}, \citenamefont {Fisher}, \citenamefont {Hirschfeld},\ and\ \citenamefont {Kotliar}}]{Fernandes2022FeSC}%
  \BibitemOpen
  \bibfield  {author} {\bibinfo {author} {\bibfnamefont {R.~M.}\ \bibnamefont {Fernandes}}, \bibinfo {author} {\bibfnamefont {A.~I.}\ \bibnamefont {Coldea}}, \bibinfo {author} {\bibfnamefont {H.}~\bibnamefont {Ding}}, \bibinfo {author} {\bibfnamefont {I.~R.}\ \bibnamefont {Fisher}}, \bibinfo {author} {\bibfnamefont {P.~J.}\ \bibnamefont {Hirschfeld}},\ and\ \bibinfo {author} {\bibfnamefont {G.}~\bibnamefont {Kotliar}},\ }\href {https://doi.org/10.1038/s41586-021-04073-2} {\bibfield  {journal} {\bibinfo  {journal} {Nature}\ }\textbf {\bibinfo {volume} {601}},\ \bibinfo {pages} {35} (\bibinfo {year} {2022})}\BibitemShut {NoStop}%
\bibitem [{\citenamefont {Aase}\ \emph {et~al.}(2023)\citenamefont {Aase}, \citenamefont {M{\ae}land},\ and\ \citenamefont {Sudb{\o}}}]{Aase2023Dec}%
  \BibitemOpen
  \bibfield  {author} {\bibinfo {author} {\bibfnamefont {N.~H.}\ \bibnamefont {Aase}}, \bibinfo {author} {\bibfnamefont {K.}~\bibnamefont {M{\ae}land}},\ and\ \bibinfo {author} {\bibfnamefont {A.}~\bibnamefont {Sudb{\o}}},\ }\href {https://doi.org/10.1103/PhysRevB.108.214508} {\bibfield  {journal} {\bibinfo  {journal} {Phys. Rev. B}\ }\textbf {\bibinfo {volume} {108}},\ \bibinfo {pages} {214508} (\bibinfo {year} {2023})}\BibitemShut {NoStop}%
\bibitem [{\citenamefont {Shang}\ \emph {et~al.}(2018)\citenamefont {Shang}, \citenamefont {Smidman}, \citenamefont {Ghosh}, \citenamefont {Baines}, \citenamefont {Chang}, \citenamefont {Gawryluk}, \citenamefont {Barker}, \citenamefont {Singh}, \citenamefont {Paul}, \citenamefont {Balakrishnan}, \citenamefont {Pomjakushina}, \citenamefont {Shi}, \citenamefont {Medarde}, \citenamefont {Hillier}, \citenamefont {Yuan}, \citenamefont {Quintanilla}, \citenamefont {Mesot},\ and\ \citenamefont {Shiroka}}]{Shang2018ReExp}%
  \BibitemOpen
  \bibfield  {author} {\bibinfo {author} {\bibfnamefont {T.}~\bibnamefont {Shang}}, \bibinfo {author} {\bibfnamefont {M.}~\bibnamefont {Smidman}}, \bibinfo {author} {\bibfnamefont {S.~K.}\ \bibnamefont {Ghosh}}, \bibinfo {author} {\bibfnamefont {C.}~\bibnamefont {Baines}}, \bibinfo {author} {\bibfnamefont {L.~J.}\ \bibnamefont {Chang}}, \bibinfo {author} {\bibfnamefont {D.~J.}\ \bibnamefont {Gawryluk}}, \bibinfo {author} {\bibfnamefont {J.~A.~T.}\ \bibnamefont {Barker}}, \bibinfo {author} {\bibfnamefont {R.~P.}\ \bibnamefont {Singh}}, \bibinfo {author} {\bibfnamefont {D.~{\relax McK}.}\ \bibnamefont {Paul}}, \bibinfo {author} {\bibfnamefont {G.}~\bibnamefont {Balakrishnan}}, \bibinfo {author} {\bibfnamefont {E.}~\bibnamefont {Pomjakushina}}, \bibinfo {author} {\bibfnamefont {M.}~\bibnamefont {Shi}}, \bibinfo {author} {\bibfnamefont {M.}~\bibnamefont {Medarde}}, \bibinfo {author} {\bibfnamefont {A.~D.}\ \bibnamefont {Hillier}}, \bibinfo {author} {\bibfnamefont {H.~Q.}\ \bibnamefont {Yuan}}, \bibinfo {author}
  {\bibfnamefont {J.}~\bibnamefont {Quintanilla}}, \bibinfo {author} {\bibfnamefont {J.}~\bibnamefont {Mesot}},\ and\ \bibinfo {author} {\bibfnamefont {T.}~\bibnamefont {Shiroka}},\ }\href {https://doi.org/10.1103/PhysRevLett.121.257002} {\bibfield  {journal} {\bibinfo  {journal} {Phys. Rev. Lett.}\ }\textbf {\bibinfo {volume} {121}},\ \bibinfo {pages} {257002} (\bibinfo {year} {2018})}\BibitemShut {NoStop}%
\bibitem [{\citenamefont {Csire}\ \emph {et~al.}(2022)\citenamefont {Csire}, \citenamefont {Annett}, \citenamefont {Quintanilla},\ and\ \citenamefont {{\ifmmode\acute{U}\else\'{U}\fi}jfalussy}}]{Csire2022ReTheory}%
  \BibitemOpen
  \bibfield  {author} {\bibinfo {author} {\bibfnamefont {G.}~\bibnamefont {Csire}}, \bibinfo {author} {\bibfnamefont {J.~F.}\ \bibnamefont {Annett}}, \bibinfo {author} {\bibfnamefont {J.}~\bibnamefont {Quintanilla}},\ and\ \bibinfo {author} {\bibfnamefont {B.}~\bibnamefont {{\ifmmode\acute{U}\else\'{U}\fi}jfalussy}},\ }\href {https://doi.org/10.1103/PhysRevB.106.L020501} {\bibfield  {journal} {\bibinfo  {journal} {Phys. Rev. B}\ }\textbf {\bibinfo {volume} {106}},\ \bibinfo {pages} {L020501} (\bibinfo {year} {2022})}\BibitemShut {NoStop}%
\bibitem [{\citenamefont {Huang}\ \emph {et~al.}(2014)\citenamefont {Huang}, \citenamefont {Taylor},\ and\ \citenamefont {Kallin}}]{Huang2014chiralpdf}%
  \BibitemOpen
  \bibfield  {author} {\bibinfo {author} {\bibfnamefont {W.}~\bibnamefont {Huang}}, \bibinfo {author} {\bibfnamefont {E.}~\bibnamefont {Taylor}},\ and\ \bibinfo {author} {\bibfnamefont {C.}~\bibnamefont {Kallin}},\ }\href {https://doi.org/10.1103/PhysRevB.90.224519} {\bibfield  {journal} {\bibinfo  {journal} {Phys. Rev. B}\ }\textbf {\bibinfo {volume} {90}},\ \bibinfo {pages} {224519} (\bibinfo {year} {2014})}\BibitemShut {NoStop}%
\bibitem [{\citenamefont {Tada}\ \emph {et~al.}(2015)\citenamefont {Tada}, \citenamefont {Nie},\ and\ \citenamefont {Oshikawa}}]{Tada2015chiralpdf}%
  \BibitemOpen
  \bibfield  {author} {\bibinfo {author} {\bibfnamefont {Y.}~\bibnamefont {Tada}}, \bibinfo {author} {\bibfnamefont {W.}~\bibnamefont {Nie}},\ and\ \bibinfo {author} {\bibfnamefont {M.}~\bibnamefont {Oshikawa}},\ }\href {https://doi.org/10.1103/PhysRevLett.114.195301} {\bibfield  {journal} {\bibinfo  {journal} {Phys. Rev. Lett.}\ }\textbf {\bibinfo {volume} {114}},\ \bibinfo {pages} {195301} (\bibinfo {year} {2015})}\BibitemShut {NoStop}%
\bibitem [{\citenamefont {Volovik}(2015)}]{Volovik2015chiralpdf}%
  \BibitemOpen
  \bibfield  {author} {\bibinfo {author} {\bibfnamefont {G.~E.}\ \bibnamefont {Volovik}},\ }\href {https://doi.org/10.1134/S0021364014230155} {\bibfield  {journal} {\bibinfo  {journal} {JETP Lett.}\ }\textbf {\bibinfo {volume} {100}},\ \bibinfo {pages} {742} (\bibinfo {year} {2015})}\BibitemShut {NoStop}%
\bibitem [{\citenamefont {Suzuki}\ and\ \citenamefont {Golubov}(2023)}]{Suzuki2023chiralpdf}%
  \BibitemOpen
  \bibfield  {author} {\bibinfo {author} {\bibfnamefont {S.-I.}\ \bibnamefont {Suzuki}}\ and\ \bibinfo {author} {\bibfnamefont {A.~A.}\ \bibnamefont {Golubov}},\ }\href {https://doi.org/10.1103/PhysRevB.108.134501} {\bibfield  {journal} {\bibinfo  {journal} {Phys. Rev. B}\ }\textbf {\bibinfo {volume} {108}},\ \bibinfo {pages} {134501} (\bibinfo {year} {2023})}\BibitemShut {NoStop}%
\bibitem [{\citenamefont {Black-Schaffer}\ and\ \citenamefont {Honerkamp}(2014)}]{Black-Schaffer2014dwave}%
  \BibitemOpen
  \bibfield  {author} {\bibinfo {author} {\bibfnamefont {A.~M.}\ \bibnamefont {Black-Schaffer}}\ and\ \bibinfo {author} {\bibfnamefont {C.}~\bibnamefont {Honerkamp}},\ }\href {https://doi.org/10.1088/0953-8984/26/42/423201} {\bibfield  {journal} {\bibinfo  {journal} {J. Phys.: Condens. Matter}\ }\textbf {\bibinfo {volume} {26}},\ \bibinfo {pages} {423201} (\bibinfo {year} {2014})}\BibitemShut {NoStop}%
\bibitem [{\citenamefont {Serban}\ \emph {et~al.}(2010)\citenamefont {Serban}, \citenamefont {B{\ifmmode\acute{e}\else\'{e}\fi}ri}, \citenamefont {Akhmerov},\ and\ \citenamefont {Beenakker}}]{Serban2010chiralpdomain}%
  \BibitemOpen
  \bibfield  {author} {\bibinfo {author} {\bibfnamefont {I.}~\bibnamefont {Serban}}, \bibinfo {author} {\bibfnamefont {B.}~\bibnamefont {B{\ifmmode\acute{e}\else\'{e}\fi}ri}}, \bibinfo {author} {\bibfnamefont {A.~R.}\ \bibnamefont {Akhmerov}},\ and\ \bibinfo {author} {\bibfnamefont {C.~W.~J.}\ \bibnamefont {Beenakker}},\ }\href {https://doi.org/10.1103/PhysRevLett.104.147001} {\bibfield  {journal} {\bibinfo  {journal} {Phys. Rev. Lett.}\ }\textbf {\bibinfo {volume} {104}},\ \bibinfo {pages} {147001} (\bibinfo {year} {2010})}\BibitemShut {NoStop}%
\bibitem [{\citenamefont {Lundemo}\ and\ \citenamefont {Sudb{\o}}(2024)}]{Lundemo2024Jan}%
  \BibitemOpen
  \bibfield  {author} {\bibinfo {author} {\bibfnamefont {S.~D.}\ \bibnamefont {Lundemo}}\ and\ \bibinfo {author} {\bibfnamefont {A.}~\bibnamefont {Sudb{\o}}},\ }\href {https://doi.org/10.1103/PhysRevB.109.184508} {\bibfield  {journal} {\bibinfo  {journal} {Phys. Rev. B}\ }\textbf {\bibinfo {volume} {109}},\ \bibinfo {pages} {184508} (\bibinfo {year} {2024})}\BibitemShut {NoStop}%
\bibitem [{\citenamefont {Schnyder}\ \emph {et~al.}(2008)\citenamefont {Schnyder}, \citenamefont {Ryu}, \citenamefont {Furusaki},\ and\ \citenamefont {Ludwig}}]{Schnyder2008tenfold}%
  \BibitemOpen
  \bibfield  {author} {\bibinfo {author} {\bibfnamefont {A.~P.}\ \bibnamefont {Schnyder}}, \bibinfo {author} {\bibfnamefont {S.}~\bibnamefont {Ryu}}, \bibinfo {author} {\bibfnamefont {A.}~\bibnamefont {Furusaki}},\ and\ \bibinfo {author} {\bibfnamefont {A.~W.~W.}\ \bibnamefont {Ludwig}},\ }\href {https://doi.org/10.1103/PhysRevB.78.195125} {\bibfield  {journal} {\bibinfo  {journal} {Phys. Rev. B}\ }\textbf {\bibinfo {volume} {78}},\ \bibinfo {pages} {195125} (\bibinfo {year} {2008})}\BibitemShut {NoStop}%
\bibitem [{\citenamefont {Bernevig}\ and\ \citenamefont {Hughes}(2013)}]{Bernevig2013}%
  \BibitemOpen
  \bibfield  {author} {\bibinfo {author} {\bibfnamefont {B.~A.}\ \bibnamefont {Bernevig}}\ and\ \bibinfo {author} {\bibfnamefont {T.~L.}\ \bibnamefont {Hughes}},\ }\href@noop {} {\emph {\bibinfo {title} {Topological Insulators and Topological Superconductors}}}\ (\bibinfo  {publisher} {Princeton University Press},\ \bibinfo {address} {Princeton, NJ},\ \bibinfo {year} {2013})\BibitemShut {NoStop}%
\bibitem [{\citenamefont {Sato}\ and\ \citenamefont {Ando}(2017)}]{TopoSCrevSato}%
  \BibitemOpen
  \bibfield  {author} {\bibinfo {author} {\bibfnamefont {M.}~\bibnamefont {Sato}}\ and\ \bibinfo {author} {\bibfnamefont {Y.}~\bibnamefont {Ando}},\ }\href {https://doi.org/10.1088/1361-6633/aa6ac7} {\bibfield  {journal} {\bibinfo  {journal} {Rep. Prog. Phys.}\ }\textbf {\bibinfo {volume} {80}},\ \bibinfo {pages} {076501} (\bibinfo {year} {2017})}\BibitemShut {NoStop}%
\bibitem [{\citenamefont {Leijnse}\ and\ \citenamefont {Flensberg}(2012)}]{Leijnse2012TSCrev}%
  \BibitemOpen
  \bibfield  {author} {\bibinfo {author} {\bibfnamefont {M.}~\bibnamefont {Leijnse}}\ and\ \bibinfo {author} {\bibfnamefont {K.}~\bibnamefont {Flensberg}},\ }\href {https://doi.org/10.1088/0268-1242/27/12/124003} {\bibfield  {journal} {\bibinfo  {journal} {Semicond. Sci. Technol.}\ }\textbf {\bibinfo {volume} {27}},\ \bibinfo {pages} {124003} (\bibinfo {year} {2012})}\BibitemShut {NoStop}%
\bibitem [{\citenamefont {Nayak}\ \emph {et~al.}(2008)\citenamefont {Nayak}, \citenamefont {Simon}, \citenamefont {Stern}, \citenamefont {Freedman},\ and\ \citenamefont {Das~Sarma}}]{TopoQuantumCompRevModPhys}%
  \BibitemOpen
  \bibfield  {author} {\bibinfo {author} {\bibfnamefont {C.}~\bibnamefont {Nayak}}, \bibinfo {author} {\bibfnamefont {S.~H.}\ \bibnamefont {Simon}}, \bibinfo {author} {\bibfnamefont {A.}~\bibnamefont {Stern}}, \bibinfo {author} {\bibfnamefont {M.}~\bibnamefont {Freedman}},\ and\ \bibinfo {author} {\bibfnamefont {S.}~\bibnamefont {Das~Sarma}},\ }\href {https://doi.org/10.1103/RevModPhys.80.1083} {\bibfield  {journal} {\bibinfo  {journal} {Rev. Mod. Phys.}\ }\textbf {\bibinfo {volume} {80}},\ \bibinfo {pages} {1083} (\bibinfo {year} {2008})}\BibitemShut {NoStop}%
\bibitem [{\citenamefont {M{\ae}land}\ \emph {et~al.}(2021)\citenamefont {M{\ae}land}, \citenamefont {R{\o}st}, \citenamefont {Wells},\ and\ \citenamefont {Sudb{\o}}}]{Maeland2021Sep}%
  \BibitemOpen
  \bibfield  {author} {\bibinfo {author} {\bibfnamefont {K.}~\bibnamefont {M{\ae}land}}, \bibinfo {author} {\bibfnamefont {H.~I.}\ \bibnamefont {R{\o}st}}, \bibinfo {author} {\bibfnamefont {J.~W.}\ \bibnamefont {Wells}},\ and\ \bibinfo {author} {\bibfnamefont {A.}~\bibnamefont {Sudb{\o}}},\ }\href {https://doi.org/10.1103/PhysRevB.104.125125} {\bibfield  {journal} {\bibinfo  {journal} {Phys. Rev. B}\ }\textbf {\bibinfo {volume} {104}},\ \bibinfo {pages} {125125} (\bibinfo {year} {2021})}\BibitemShut {NoStop}%
\bibitem [{\citenamefont {M{\ae}land}\ and\ \citenamefont {Sudb{\o}}(2023{\natexlab{b}})}]{Maeland2023EliashCC}%
  \BibitemOpen
  \bibfield  {author} {\bibinfo {author} {\bibfnamefont {K.}~\bibnamefont {M{\ae}land}}\ and\ \bibinfo {author} {\bibfnamefont {A.}~\bibnamefont {Sudb{\o}}},\ }\href {https://doi.org/10.1103/PhysRevB.108.214511} {\bibfield  {journal} {\bibinfo  {journal} {Phys. Rev. B}\ }\textbf {\bibinfo {volume} {108}},\ \bibinfo {pages} {214511} (\bibinfo {year} {2023}{\natexlab{b}})}\BibitemShut {NoStop}%
\bibitem [{\citenamefont {Vidberg}\ and\ \citenamefont {Serene}(1977)}]{Vidberg1977Nov}%
  \BibitemOpen
  \bibfield  {author} {\bibinfo {author} {\bibfnamefont {H.~J.}\ \bibnamefont {Vidberg}}\ and\ \bibinfo {author} {\bibfnamefont {J.~W.}\ \bibnamefont {Serene}},\ }\href {https://doi.org/10.1007/BF00655090} {\bibfield  {journal} {\bibinfo  {journal} {J. Low Temp. Phys.}\ }\textbf {\bibinfo {volume} {29}},\ \bibinfo {pages} {179} (\bibinfo {year} {1977})}\BibitemShut {NoStop}%
\bibitem [{\citenamefont {Marsiglio}\ \emph {et~al.}(1988)\citenamefont {Marsiglio}, \citenamefont {Schossmann},\ and\ \citenamefont {Carbotte}}]{Marsiglio1988Apr}%
  \BibitemOpen
  \bibfield  {author} {\bibinfo {author} {\bibfnamefont {F.}~\bibnamefont {Marsiglio}}, \bibinfo {author} {\bibfnamefont {M.}~\bibnamefont {Schossmann}},\ and\ \bibinfo {author} {\bibfnamefont {J.~P.}\ \bibnamefont {Carbotte}},\ }\href {https://doi.org/10.1103/PhysRevB.37.4965} {\bibfield  {journal} {\bibinfo  {journal} {Phys. Rev. B}\ }\textbf {\bibinfo {volume} {37}},\ \bibinfo {pages} {4965} (\bibinfo {year} {1988})}\BibitemShut {NoStop}%
\bibitem [{\citenamefont {Aperis}\ \emph {et~al.}(2015)\citenamefont {Aperis}, \citenamefont {Maldonado},\ and\ \citenamefont {Oppeneer}}]{Aperis2015Aug}%
  \BibitemOpen
  \bibfield  {author} {\bibinfo {author} {\bibfnamefont {A.}~\bibnamefont {Aperis}}, \bibinfo {author} {\bibfnamefont {P.}~\bibnamefont {Maldonado}},\ and\ \bibinfo {author} {\bibfnamefont {P.~M.}\ \bibnamefont {Oppeneer}},\ }\href {https://doi.org/10.1103/PhysRevB.92.054516} {\bibfield  {journal} {\bibinfo  {journal} {Phys. Rev. B}\ }\textbf {\bibinfo {volume} {92}},\ \bibinfo {pages} {054516} (\bibinfo {year} {2015})}\BibitemShut {NoStop}%
\bibitem [{\citenamefont {Aperis}\ and\ \citenamefont {Oppeneer}(2018)}]{Aperis2018Feb}%
  \BibitemOpen
  \bibfield  {author} {\bibinfo {author} {\bibfnamefont {A.}~\bibnamefont {Aperis}}\ and\ \bibinfo {author} {\bibfnamefont {P.~M.}\ \bibnamefont {Oppeneer}},\ }\href {https://doi.org/10.1103/PhysRevB.97.060501} {\bibfield  {journal} {\bibinfo  {journal} {Phys. Rev. B}\ }\textbf {\bibinfo {volume} {97}},\ \bibinfo {pages} {060501} (\bibinfo {year} {2018})}\BibitemShut {NoStop}%
\bibitem [{\citenamefont {Schrodi}\ \emph {et~al.}(2020{\natexlab{a}})\citenamefont {Schrodi}, \citenamefont {Aperis},\ and\ \citenamefont {Oppeneer}}]{Schrodi2020Mar}%
  \BibitemOpen
  \bibfield  {author} {\bibinfo {author} {\bibfnamefont {F.}~\bibnamefont {Schrodi}}, \bibinfo {author} {\bibfnamefont {A.}~\bibnamefont {Aperis}},\ and\ \bibinfo {author} {\bibfnamefont {P.~M.}\ \bibnamefont {Oppeneer}},\ }\href {https://doi.org/10.1103/PhysRevResearch.2.012066} {\bibfield  {journal} {\bibinfo  {journal} {Phys. Rev. Res.}\ }\textbf {\bibinfo {volume} {2}},\ \bibinfo {pages} {012066} (\bibinfo {year} {2020}{\natexlab{a}})}\BibitemShut {NoStop}%
\bibitem [{\citenamefont {Schrodi}\ \emph {et~al.}(2020{\natexlab{b}})\citenamefont {Schrodi}, \citenamefont {Oppeneer},\ and\ \citenamefont {Aperis}}]{Schrodi2020FullBandwidth}%
  \BibitemOpen
  \bibfield  {author} {\bibinfo {author} {\bibfnamefont {F.}~\bibnamefont {Schrodi}}, \bibinfo {author} {\bibfnamefont {P.~M.}\ \bibnamefont {Oppeneer}},\ and\ \bibinfo {author} {\bibfnamefont {A.}~\bibnamefont {Aperis}},\ }\href {https://doi.org/10.1103/PhysRevB.102.024503} {\bibfield  {journal} {\bibinfo  {journal} {Phys. Rev. B}\ }\textbf {\bibinfo {volume} {102}},\ \bibinfo {pages} {024503} (\bibinfo {year} {2020}{\natexlab{b}})}\BibitemShut {NoStop}%
\bibitem [{\citenamefont {Senthil}\ \emph {et~al.}(1999)\citenamefont {Senthil}, \citenamefont {Marston},\ and\ \citenamefont {Fisher}}]{Senthil1999dwave}%
  \BibitemOpen
  \bibfield  {author} {\bibinfo {author} {\bibfnamefont {T.}~\bibnamefont {Senthil}}, \bibinfo {author} {\bibfnamefont {J.~B.}\ \bibnamefont {Marston}},\ and\ \bibinfo {author} {\bibfnamefont {M.~P.~A.}\ \bibnamefont {Fisher}},\ }\href {https://doi.org/10.1103/PhysRevB.60.4245} {\bibfield  {journal} {\bibinfo  {journal} {Phys. Rev. B}\ }\textbf {\bibinfo {volume} {60}},\ \bibinfo {pages} {4245} (\bibinfo {year} {1999})}\BibitemShut {NoStop}%
\bibitem [{\citenamefont {Vojta}\ \emph {et~al.}(2000)\citenamefont {Vojta}, \citenamefont {Zhang},\ and\ \citenamefont {Sachdev}}]{Vojta2000dwave}%
  \BibitemOpen
  \bibfield  {author} {\bibinfo {author} {\bibfnamefont {M.}~\bibnamefont {Vojta}}, \bibinfo {author} {\bibfnamefont {Y.}~\bibnamefont {Zhang}},\ and\ \bibinfo {author} {\bibfnamefont {S.}~\bibnamefont {Sachdev}},\ }\href {https://doi.org/10.1103/PhysRevLett.85.4940} {\bibfield  {journal} {\bibinfo  {journal} {Phys. Rev. Lett.}\ }\textbf {\bibinfo {volume} {85}},\ \bibinfo {pages} {4940} (\bibinfo {year} {2000})}\BibitemShut {NoStop}%
\bibitem [{\citenamefont {Black-Schaffer}(2012)}]{Black-Schaffer2012dwave}%
  \BibitemOpen
  \bibfield  {author} {\bibinfo {author} {\bibfnamefont {A.~M.}\ \bibnamefont {Black-Schaffer}},\ }\href {https://doi.org/10.1103/PhysRevLett.109.197001} {\bibfield  {journal} {\bibinfo  {journal} {Phys. Rev. Lett.}\ }\textbf {\bibinfo {volume} {109}},\ \bibinfo {pages} {197001} (\bibinfo {year} {2012})}\BibitemShut {NoStop}%
\bibitem [{\citenamefont {Fischer}\ \emph {et~al.}(2014)\citenamefont {Fischer}, \citenamefont {Neupert}, \citenamefont {Platt}, \citenamefont {Schnyder}, \citenamefont {Hanke}, \citenamefont {Goryo}, \citenamefont {Thomale},\ and\ \citenamefont {Sigrist}}]{Fischer2014dwave}%
  \BibitemOpen
  \bibfield  {author} {\bibinfo {author} {\bibfnamefont {M.~H.}\ \bibnamefont {Fischer}}, \bibinfo {author} {\bibfnamefont {T.}~\bibnamefont {Neupert}}, \bibinfo {author} {\bibfnamefont {C.}~\bibnamefont {Platt}}, \bibinfo {author} {\bibfnamefont {A.~P.}\ \bibnamefont {Schnyder}}, \bibinfo {author} {\bibfnamefont {W.}~\bibnamefont {Hanke}}, \bibinfo {author} {\bibfnamefont {J.}~\bibnamefont {Goryo}}, \bibinfo {author} {\bibfnamefont {R.}~\bibnamefont {Thomale}},\ and\ \bibinfo {author} {\bibfnamefont {M.}~\bibnamefont {Sigrist}},\ }\href {https://doi.org/10.1103/PhysRevB.89.020509} {\bibfield  {journal} {\bibinfo  {journal} {Phys. Rev. B}\ }\textbf {\bibinfo {volume} {89}},\ \bibinfo {pages} {020509} (\bibinfo {year} {2014})}\BibitemShut {NoStop}%
\bibitem [{\citenamefont {Holmvall}\ and\ \citenamefont {Black-Schaffer}(2023)}]{Holmvall2023dwave}%
  \BibitemOpen
  \bibfield  {author} {\bibinfo {author} {\bibfnamefont {P.}~\bibnamefont {Holmvall}}\ and\ \bibinfo {author} {\bibfnamefont {A.~M.}\ \bibnamefont {Black-Schaffer}},\ }\href {https://doi.org/10.1103/PhysRevB.108.L100506} {\bibfield  {journal} {\bibinfo  {journal} {Phys. Rev. B}\ }\textbf {\bibinfo {volume} {108}},\ \bibinfo {pages} {L100506} (\bibinfo {year} {2023})}\BibitemShut {NoStop}%
\bibitem [{\citenamefont {Kitaev}(2001)}]{Kitaev2001Oct}%
  \BibitemOpen
  \bibfield  {author} {\bibinfo {author} {\bibfnamefont {A.~Y.}\ \bibnamefont {Kitaev}},\ }\href {https://doi.org/10.1070/1063-7869/44/10S/S29} {\bibfield  {journal} {\bibinfo  {journal} {Phys.-Usp.}\ }\textbf {\bibinfo {volume} {44}},\ \bibinfo {pages} {131} (\bibinfo {year} {2001})}\BibitemShut {NoStop}%
\bibitem [{\citenamefont {Oreg}\ \emph {et~al.}(2010)\citenamefont {Oreg}, \citenamefont {Refael},\ and\ \citenamefont {von Oppen}}]{Oreg2010Oct}%
  \BibitemOpen
  \bibfield  {author} {\bibinfo {author} {\bibfnamefont {Y.}~\bibnamefont {Oreg}}, \bibinfo {author} {\bibfnamefont {G.}~\bibnamefont {Refael}},\ and\ \bibinfo {author} {\bibfnamefont {F.}~\bibnamefont {von Oppen}},\ }\href {https://doi.org/10.1103/PhysRevLett.105.177002} {\bibfield  {journal} {\bibinfo  {journal} {Phys. Rev. Lett.}\ }\textbf {\bibinfo {volume} {105}},\ \bibinfo {pages} {177002} (\bibinfo {year} {2010})}\BibitemShut {NoStop}%
\bibitem [{\citenamefont {{M. Aghaee \textit{et al.} (Microsoft Quantum)}}(2023)}]{MicrosoftTGP}%
  \BibitemOpen
  \bibfield  {author} {\bibinfo {author} {\bibnamefont {{M. Aghaee \textit{et al.} (Microsoft Quantum)}}},\ }\href {https://doi.org/10.1103/PhysRevB.107.245423} {\bibfield  {journal} {\bibinfo  {journal} {Phys. Rev. B}\ }\textbf {\bibinfo {volume} {107}},\ \bibinfo {pages} {245423} (\bibinfo {year} {2023})}\BibitemShut {NoStop}%
\bibitem [{\citenamefont {Hell}\ \emph {et~al.}(2017)\citenamefont {Hell}, \citenamefont {Leijnse},\ and\ \citenamefont {Flensberg}}]{Hell2017JJTSC}%
  \BibitemOpen
  \bibfield  {author} {\bibinfo {author} {\bibfnamefont {M.}~\bibnamefont {Hell}}, \bibinfo {author} {\bibfnamefont {M.}~\bibnamefont {Leijnse}},\ and\ \bibinfo {author} {\bibfnamefont {K.}~\bibnamefont {Flensberg}},\ }\href {https://doi.org/10.1103/PhysRevLett.118.107701} {\bibfield  {journal} {\bibinfo  {journal} {Phys. Rev. Lett.}\ }\textbf {\bibinfo {volume} {118}},\ \bibinfo {pages} {107701} (\bibinfo {year} {2017})}\BibitemShut {NoStop}%
\bibitem [{\citenamefont {Pientka}\ \emph {et~al.}(2017)\citenamefont {Pientka}, \citenamefont {Keselman}, \citenamefont {Berg}, \citenamefont {Yacoby}, \citenamefont {Stern},\ and\ \citenamefont {Halperin}}]{Pientka2017JJTSC}%
  \BibitemOpen
  \bibfield  {author} {\bibinfo {author} {\bibfnamefont {F.}~\bibnamefont {Pientka}}, \bibinfo {author} {\bibfnamefont {A.}~\bibnamefont {Keselman}}, \bibinfo {author} {\bibfnamefont {E.}~\bibnamefont {Berg}}, \bibinfo {author} {\bibfnamefont {A.}~\bibnamefont {Yacoby}}, \bibinfo {author} {\bibfnamefont {A.}~\bibnamefont {Stern}},\ and\ \bibinfo {author} {\bibfnamefont {B.~I.}\ \bibnamefont {Halperin}},\ }\href {https://doi.org/10.1103/PhysRevX.7.021032} {\bibfield  {journal} {\bibinfo  {journal} {Phys. Rev. X}\ }\textbf {\bibinfo {volume} {7}},\ \bibinfo {pages} {021032} (\bibinfo {year} {2017})}\BibitemShut {NoStop}%
\bibitem [{\citenamefont {Lesser}\ \emph {et~al.}(2022)\citenamefont {Lesser}, \citenamefont {Oreg},\ and\ \citenamefont {Stern}}]{Lesser2022JJTSC}%
  \BibitemOpen
  \bibfield  {author} {\bibinfo {author} {\bibfnamefont {O.}~\bibnamefont {Lesser}}, \bibinfo {author} {\bibfnamefont {Y.}~\bibnamefont {Oreg}},\ and\ \bibinfo {author} {\bibfnamefont {A.}~\bibnamefont {Stern}},\ }\href {https://doi.org/10.1103/PhysRevB.106.L241405} {\bibfield  {journal} {\bibinfo  {journal} {Phys. Rev. B}\ }\textbf {\bibinfo {volume} {106}},\ \bibinfo {pages} {L241405} (\bibinfo {year} {2022})}\BibitemShut {NoStop}%
\bibitem [{\citenamefont {Zlotnikov}\ \emph {et~al.}(2021)\citenamefont {Zlotnikov}, \citenamefont {Shustin},\ and\ \citenamefont {Fedoseev}}]{TopoSCandSkRev}%
  \BibitemOpen
  \bibfield  {author} {\bibinfo {author} {\bibfnamefont {A.~O.}\ \bibnamefont {Zlotnikov}}, \bibinfo {author} {\bibfnamefont {M.~S.}\ \bibnamefont {Shustin}},\ and\ \bibinfo {author} {\bibfnamefont {A.~D.}\ \bibnamefont {Fedoseev}},\ }\href {https://doi.org/10.1007/s10948-021-06029-z} {\bibfield  {journal} {\bibinfo  {journal} {J. Supercond. Nov. Magn.}\ }\textbf {\bibinfo {volume} {34}},\ \bibinfo {pages} {3053} (\bibinfo {year} {2021})}\BibitemShut {NoStop}%
\bibitem [{\citenamefont {Hasan}\ and\ \citenamefont {Kane}(2010)}]{Hasan2010BulkTop}%
  \BibitemOpen
  \bibfield  {author} {\bibinfo {author} {\bibfnamefont {M.~Z.}\ \bibnamefont {Hasan}}\ and\ \bibinfo {author} {\bibfnamefont {C.~L.}\ \bibnamefont {Kane}},\ }\href {https://doi.org/10.1103/RevModPhys.82.3045} {\bibfield  {journal} {\bibinfo  {journal} {Rev. Mod. Phys.}\ }\textbf {\bibinfo {volume} {82}},\ \bibinfo {pages} {3045} (\bibinfo {year} {2010})}\BibitemShut {NoStop}%
\bibitem [{\citenamefont {Sato}\ \emph {et~al.}(2014)\citenamefont {Sato}, \citenamefont {Yamakage},\ and\ \citenamefont {Mizushima}}]{Sato2014BulkTopSpinful}%
  \BibitemOpen
  \bibfield  {author} {\bibinfo {author} {\bibfnamefont {M.}~\bibnamefont {Sato}}, \bibinfo {author} {\bibfnamefont {A.}~\bibnamefont {Yamakage}},\ and\ \bibinfo {author} {\bibfnamefont {T.}~\bibnamefont {Mizushima}},\ }\href {https://doi.org/10.1016/j.physe.2013.07.011} {\bibfield  {journal} {\bibinfo  {journal} {Physica E}\ }\textbf {\bibinfo {volume} {55}},\ \bibinfo {pages} {20} (\bibinfo {year} {2014})}\BibitemShut {NoStop}%
\bibitem [{\citenamefont {Mochol-Grzelak}\ \emph {et~al.}(2018)\citenamefont {Mochol-Grzelak}, \citenamefont {Dauphin}, \citenamefont {Celi},\ and\ \citenamefont {Lewenstein}}]{Mochol-Grzelak2018BulkTopDegen}%
  \BibitemOpen
  \bibfield  {author} {\bibinfo {author} {\bibfnamefont {M.}~\bibnamefont {Mochol-Grzelak}}, \bibinfo {author} {\bibfnamefont {A.}~\bibnamefont {Dauphin}}, \bibinfo {author} {\bibfnamefont {A.}~\bibnamefont {Celi}},\ and\ \bibinfo {author} {\bibfnamefont {M.}~\bibnamefont {Lewenstein}},\ }\href {https://doi.org/10.1088/2058-9565/aae93b} {\bibfield  {journal} {\bibinfo  {journal} {Quantum Sci. Technol.}\ }\textbf {\bibinfo {volume} {4}},\ \bibinfo {pages} {014009} (\bibinfo {year} {2018})}\BibitemShut {NoStop}%
\bibitem [{\citenamefont {Sachdev}(2023)}]{sachdev2023quantum}%
  \BibitemOpen
  \bibfield  {author} {\bibinfo {author} {\bibfnamefont {S.}~\bibnamefont {Sachdev}},\ }\href@noop {} {\emph {\bibinfo {title} {Quantum Phases of Matter}}}\ (\bibinfo  {publisher} {Cambridge University Press},\ \bibinfo {address} {Cambridge, UK},\ \bibinfo {year} {2023})\BibitemShut {NoStop}%
\bibitem [{\citenamefont {Altland}\ and\ \citenamefont {Zirnbauer}(1997)}]{Altland1997tenfold}%
  \BibitemOpen
  \bibfield  {author} {\bibinfo {author} {\bibfnamefont {A.}~\bibnamefont {Altland}}\ and\ \bibinfo {author} {\bibfnamefont {M.~R.}\ \bibnamefont {Zirnbauer}},\ }\href {https://doi.org/10.1103/PhysRevB.55.1142} {\bibfield  {journal} {\bibinfo  {journal} {Phys. Rev. B}\ }\textbf {\bibinfo {volume} {55}},\ \bibinfo {pages} {1142} (\bibinfo {year} {1997})}\BibitemShut {NoStop}%
\bibitem [{\citenamefont {Teo}\ and\ \citenamefont {Kane}(2010)}]{TeoKane2010defectdm1}%
  \BibitemOpen
  \bibfield  {author} {\bibinfo {author} {\bibfnamefont {J.~C.~Y.}\ \bibnamefont {Teo}}\ and\ \bibinfo {author} {\bibfnamefont {C.~L.}\ \bibnamefont {Kane}},\ }\href {https://doi.org/10.1103/PhysRevB.82.115120} {\bibfield  {journal} {\bibinfo  {journal} {Phys. Rev. B}\ }\textbf {\bibinfo {volume} {82}},\ \bibinfo {pages} {115120} (\bibinfo {year} {2010})}\BibitemShut {NoStop}%
\bibitem [{\citenamefont {Mercado}\ \emph {et~al.}(2022)\citenamefont {Mercado}, \citenamefont {Sahoo},\ and\ \citenamefont {Franz}}]{Mercado2022Majoranadwave}%
  \BibitemOpen
  \bibfield  {author} {\bibinfo {author} {\bibfnamefont {A.}~\bibnamefont {Mercado}}, \bibinfo {author} {\bibfnamefont {S.}~\bibnamefont {Sahoo}},\ and\ \bibinfo {author} {\bibfnamefont {M.}~\bibnamefont {Franz}},\ }\href {https://doi.org/10.1103/PhysRevLett.128.137002} {\bibfield  {journal} {\bibinfo  {journal} {Phys. Rev. Lett.}\ }\textbf {\bibinfo {volume} {128}},\ \bibinfo {pages} {137002} (\bibinfo {year} {2022})}\BibitemShut {NoStop}%
\bibitem [{\citenamefont {Margalit}\ \emph {et~al.}(2022)\citenamefont {Margalit}, \citenamefont {Yan}, \citenamefont {Franz},\ and\ \citenamefont {Oreg}}]{Margalit2022Majoranadwave}%
  \BibitemOpen
  \bibfield  {author} {\bibinfo {author} {\bibfnamefont {G.}~\bibnamefont {Margalit}}, \bibinfo {author} {\bibfnamefont {B.}~\bibnamefont {Yan}}, \bibinfo {author} {\bibfnamefont {M.}~\bibnamefont {Franz}},\ and\ \bibinfo {author} {\bibfnamefont {Y.}~\bibnamefont {Oreg}},\ }\href {https://doi.org/10.1103/PhysRevB.106.205424} {\bibfield  {journal} {\bibinfo  {journal} {Phys. Rev. B}\ }\textbf {\bibinfo {volume} {106}},\ \bibinfo {pages} {205424} (\bibinfo {year} {2022})}\BibitemShut {NoStop}%
\end{thebibliography}%
\end{document}